\documentclass[preprint,12pt]{elsarticle}

\usepackage{amssymb}

\journal{Journal of Quantitative Spectroscopy and Radiative Transfer}
\usepackage[hyphens]{url}
\usepackage{kbordermatrix}
\renewcommand{\kbldelim}{(} 
\renewcommand{\kbrdelim}{)}
\usepackage{pgfplots}
\usepackage[utf8]{inputenc}
\pgfplotsset{compat=1.14}
\usepackage{tikz}
\usepackage{amssymb}
\usepackage{circuitikz}
\usetikzlibrary{shapes,positioning} 
\usetikzlibrary{decorations.markings,decorations.pathmorphing,decorations.pathreplacing}
\usetikzlibrary{patterns,shapes.geometric} 
\usetikzlibrary{shapes.multipart}
\usetikzlibrary{arrows}
\tikzstyle directed=[postaction={decorate,decoration={markings,
		mark=at position .65 with {\arrow{latex}}}}]
\usepackage{multirow}
\usepackage{mathtools}
\usepackage{braket}
\usepackage{lipsum}
\usepackage{amsmath}
\biboptions{numbers,sort&compress}
\usepackage{float}

\begin{document}

\begin{frontmatter}



\title{New standard magnetic field values determined by cancellations of ${}^{85}\text{Rb}$ and ${}^{87}\text{Rb}$ atomic vapors $5\prescript{2}{}{S}_{1/2} \rightarrow 6\prescript{2}{}{P}_{1/2,~3/2}$ transitions}

 \author[label1]{Rodolphe MOMIER}
  \author[label1,label2]{Artur ALEKSANYAN}
  \author[label2,label3]{Emil GAZAZYAN}
   \author[label2]{Aram PAPOYAN}
   \author[label1]{Claude LEROY\corref{cor1}}
   
    \cortext[cor1]{Corresponding author: claude.leroy@u-bourgogne.fr}

\address[label1]{Laboratoire Interdisciplinaire Carnot de Bourgogne, UMR CNRS 6303, Université Bourgogne Franche-Comté, 21000 Dijon, France}
\address[label2]{Institute for Physical Research, NAS of Armenia, Ashtarak-2, 0203 Armenia}
\address[label3]{Yerevan State University, Yerevan, 0025 Armenia}

\address{}

\begin{graphicalabstract}
\includegraphics[width=\textwidth]{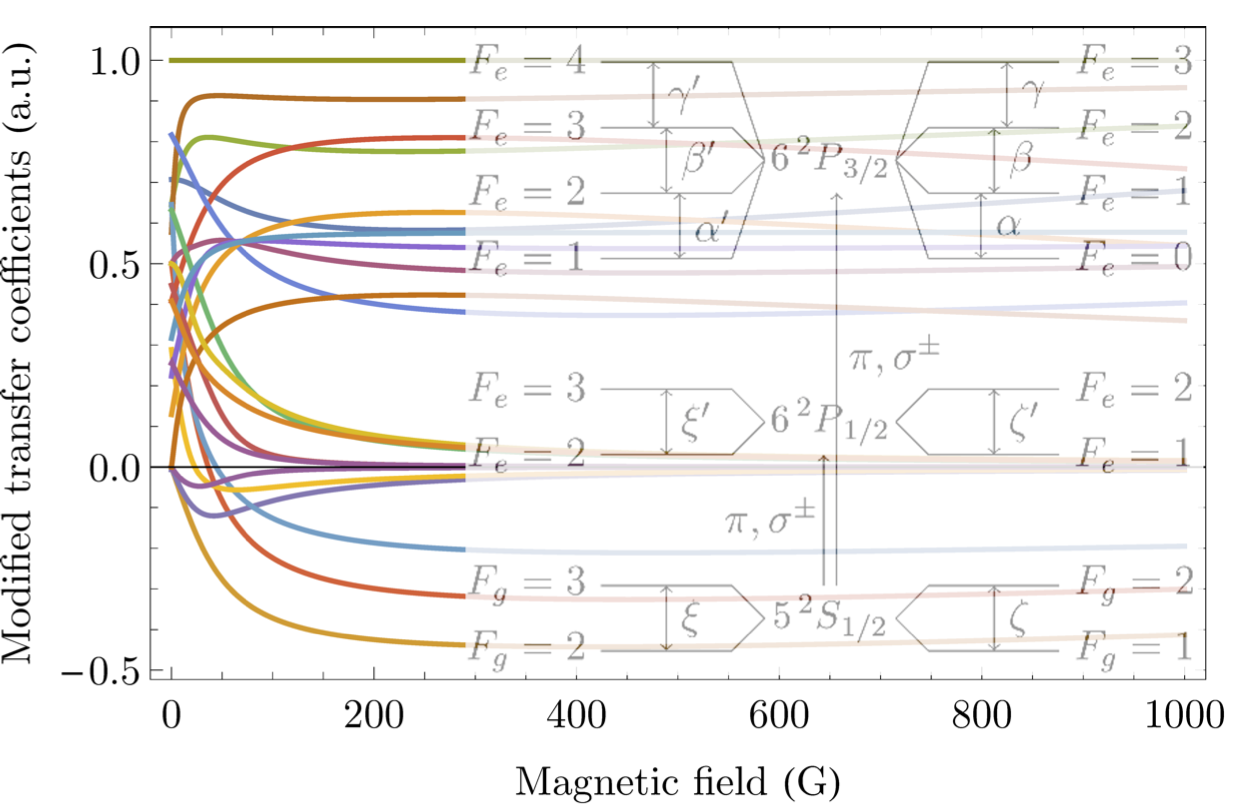}
\end{graphicalabstract}

\begin{abstract}
In this article, we study the theoretical behaviour of all the possible hyperfine transitions ($\pi$, $\sigma^+$ and $\sigma^-$) between the $5S$ and $6P$ states of ${}^{87}\text{Rb}$ and ${}^{85}\text{Rb}$ atomic vapors under the influence of an external magnetic field $B$. We show that, for specific transitions, we obtain one or several $B$-values for which the transition intensity is cancelled. The precision of these values is limited to the uncertainty of the physical quantities that are involved in the problem, thus measuring precisely the $B$-values for the cancellations could be a way to determine these quantities more precisely. In the simplest cases involving $2\times 2$ hamiltonians, we give eigenvectors, eigenvalues and analytical formulas to determine the transition cancellation. By checking accuracy between formulas and numerical simulations, we conclude that it is possible to use the latter in order to determine all the cancellations even in the most complicated cases.
\end{abstract}

\begin{highlights}
\item The intensities of several hyperfine transitions of an atomic vapor  in an external magnetic field are cancelled for very precise values of the magnetic field.

\item It is now possible to determine transition cancellations numerically and, in some cases, analytically.

\item Measuring (precisely) the magnetic field values could help refining the values of some physical constants. 
\end{highlights}


\end{frontmatter}


\section{Introduction}

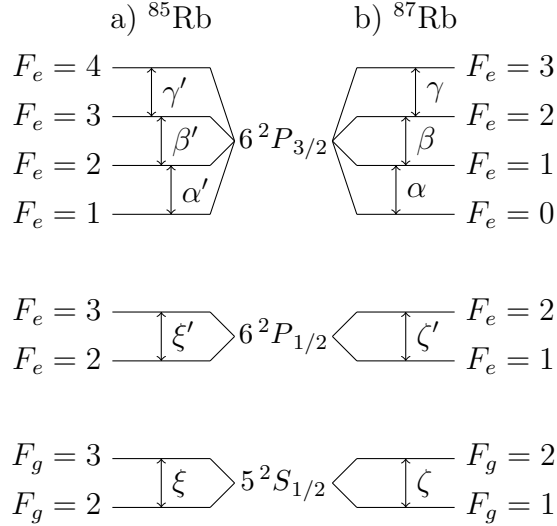
\begin{figure}[H]
\centering
\begin{tikzpicture} [scale=0.65, decoration={coil,aspect=0.4,segment length=3mm,amplitude=3mm}]

\draw (7,0) -- (9,0) node [right] {$F_g = 1$} ;
\draw(7,1) -- (9,1) node [right] {$F_g = 2$} ;
\draw (7,0) -- (6.5,0.5);
\draw (7,1) -- (6.5,0.5);

\draw(7,3) -- (9,3) node [right] {$F_e = 1$} ;
\draw(7,4) -- (9,4) node [right] {$F_e = 2$} ;
\draw (7,3) -- (6.5,3.5);
\draw (7,4) -- (6.5,3.5);

\draw (2,0) -- (4,0) node [left] at (2,0) {$F_g = 2$} ;
\draw(2,1) -- (4,1) node [left] at (2,1){$F_g = 3$} ;
\draw (4,0) -- (4.5,0.5);
\draw (4,1) -- (4.5,0.5);

\draw (2,3) -- (4,3) node [left] at (2,3) {$F_e = 2$} ;
\draw(2,4) -- (4,4) node [left] at (2,4){$F_e = 3$} ;
\draw (4,3) -- (4.5,3.5);
\draw (4,4) -- (4.5,3.5);

\draw(7,6) -- (9,6) node [right] {$F_e = 0$} ;
\draw(7,7) -- (9,7) node [right] {$F_e = 1$} ; 
\draw(7,8) -- (9,8) node [right] {$F_e = 2$} ;
\draw(7,9) -- (9,9) node [right] {$F_e = 3$} ;
\draw(7,6) -- (6.5,7.5);
\draw(7,7) -- (6.5,7.5);
\draw(7,8) -- (6.5,7.5);
\draw(7,9) -- (6.5,7.5);

\draw(2,6) -- (4,6) node [left]  at (2,6) {$F_e = 1$} ;
\draw(2,7) -- (4,7) node [left] at (2,7) {$F_e = 2$} ; 
\draw(2,8) -- (4,8) node [left] at (2,8) {$F_e = 3$} ;
\draw(2,9) -- (4,9) node [left] at (2,9) {$F_e = 4$} ;
\draw(4,6) -- (4.5,7.5);
\draw(4,7) -- (4.5,7.5);
\draw(4,8) -- (4.5,7.5);
\draw(4,9) -- (4.5,7.5);

\node at (5.5,0.5)  {\small $5\prescript{2}{}{S}_{1/2}$};
\node at (5.5,3.5)  {\small $6\prescript{2}{}{P}_{1/2}$};
\node at (5.5,7.5)  {\small $6\prescript{2}{}{P}_{3/2}$};


\draw[<->](8,0) -- node[right] {$\zeta$} (8,1) ;
\draw[<->](8,3) -- node[right] {$\zeta '$} (8,4) ;
\draw[<->](7.8,6) -- node[right] {$\alpha$} (7.8,7) ;
\draw[<->](8,7) -- node[right] {$\beta$} (8,8) ;
\draw[<->](8.2,8) -- node[right] {$\gamma$} (8.2,9) ;

\draw[<->](3,0) -- node[right] {$\xi$} (3,1) ;
\draw[<->](3,3) -- node[right] {$\xi '$} (3,4) ;
\draw[<->](3.2,6) -- node[right] {$\alpha '$} (3.2,7) ;
\draw[<->](3,7) -- node[right] {$\beta '$} (3,8) ;
\draw[<->](2.8,8) -- node[right] {$\gamma '$} (2.8,9) ;

\node at (3,10) {a) ${}^{85}\text{Rb}$};
\node at (8,10) {b) ${}^{87}\text{Rb}$};
\end{tikzpicture}
%
%
%
\caption{Hyperfine manifold of the $5S$ and $6P$ states of a) ${}^{85}\text{Rb}$ and b) ${}^{87}\text{Rb}$. Figure not to scale. See table \ref{tab:hf_splittings} for the splittings. \label{fig:hyperfine_str}}
\end{figure}

Alkali have been deeply studied for several years mainly because of their one-electron structure in their valence shell. Spectroscopic data for these metals are well known \cite{RevModPhys.49.31}. Due to their simple energy-level structure, alkali are widely used to study Bose-Einstein condensates \cite{bec}. Moreover, alkali have transition frequencies ranging from near-infrared to mid-visible, thus is it possible to do experiments with cheap lasers. 
Alkali are also, for example, studied in magnetometry or information storage \cite{Legaie:18,infstore}.
When placed under the influence of an external magnetic field $B$, energy levels split into hyperfine sublevels (Zeeman effect)      
\cite{OPA,so,Weller_2012}. The hyperfine manifold of 5S and 6P states of both rubidium isotopes is presented on figure \ref{fig:hyperfine_str}.  The theoretical model to study transitions between these hyperfine sublevels has been developed in the 1990s by Tremblay \textit{et~al.} \cite{PhysRevA.42.2766}. 
It is well known that in intermediate magnetic fields, the splitting of atomic energy levels into Zeeman sublevels deviates from the linear behavior, and the atomic transition probabilities undergo significant changes \cite{PhysRevA.42.2766,spectro}.
To study such transitions, sub-Doppler methods have to be used because of Doppler broadening \cite{khan}.
 A lot of work has been performed concerning $D_1$ and $D_2$ lines and good results have been obtained using derivative selective reflection method \cite{dijonash1,dijonash2,dijonash3}.

In this article, we study $5\prescript{2}{}{S}_{1/2} \rightarrow 6\prescript{2}{}{P}_{1/2,~3/2}$ transitions. Very little information on these transitions is available in the literature \cite{glaser}. The energy level manifold is presented in figure \ref{fig:hyperfine_str} and hyperfine splittings (along with references) are provided in table \ref{tab:hf_splittings}. We will compute all the possible transitions between Zeeman sublevels and present a full formalism for the most simple cases ($5\prescript{2}{}{S}_{1/2} \rightarrow 6\prescript{2}{}{P}_{1/2}$ transitions) to determine transition cancellations. For the case $5\prescript{2}{}{S}_{1/2} \rightarrow 6\prescript{2}{}{P}_{3/2}$, we present numerical results.

\begin{table}[H]
\centering
\begin{tabular}{|c|c|c|c|}
\hline
 
Atom                                 & State                                         & Hyperfine splitting (MHz)       & \multicolumn{1}{c|}{Reference} \\ \hline \hline
                                     & $5\prescript{2}{}{S}_{1/2}$                   & $\xi = 3035.7324390(60)$        &                                                       \cite{steck85} \\ \cline{2-4} 
                                     & $6\prescript{2}{}{P}_{1/2}$                   & $\xi ' = 118.40(46)$            &                                                        \\ \cline{2-3}
                                     &                                               & $\alpha '= 10.048(25)$          &                                                        \\ \cline{3-3}
                                     &                                               & $\beta ' = 20.967(25)$          &                                                        \\ \cline{3-3}
\multirow{-5}{*}{${}^{85}\text{Rb}$} & \multirow{-3}{*}{$6\prescript{2}{}{P}_{3/2}$} & $\gamma ' = 39.127(25)$         & \multirow{-4}{*}{\cite{glaser}}                                     \\ \hline \hline
                                     & $5\prescript{2}{}{S}_{1/2}$                   & $\zeta = 6834.682610904290(90)$ &                                                       \cite{steck87} \\ \cline{2-4} 
                                     & $6\prescript{2}{}{P}_{1/2}$                   & $\zeta ' = 265.15(46)$          &                                                        \\ \cline{2-3}
                                     &                                               & $\alpha = 23.739(26)$           &                                                        \\ \cline{3-3}
                                     &                                               & $\beta = 51.654(26)$            &                                                        \\ \cline{3-3}
\multirow{-5}{*}{${}^{87}\text{Rb}$} & \multirow{-3}{*}{$6\prescript{2}{}{P}_{3/2}$} & $\gamma = 87.009(26)$           & \multirow{-4}{*}{\cite{glaser}}                                     \\ \hline
\end{tabular}
\caption{Numerical values of the $5S$ and $6P$ hyperfine splittings.}
\label{tab:hf_splittings}
\end{table}

\section{Theoretical model}

\subsection{Hamiltonian of an atomic alkali vapor under the influence of an external magnetic field}
The Hamiltonian $\mathcal{H}$ of an atomic alkali vapor placed under the influence of an external static magnetic field $B$ can be computed using the formalism presented in \cite{PhysRevA.42.2766}. In the basis of the unperturbated atomic state vectors $\ket{F,m_F}$, matrix elements are given by

\begin{equation}
\bra{F,m_F}\mathcal{H}\ket{F,m_F} = E_0(F) - \mu_Bg_Fm_FB \label{eq:diag}
\end{equation}
and
\begin{align}
& \bra{F-1,m_F}\mathcal{H}\ket{F,m_F} = \bra{F,m_F}\mathcal{H}\ket{F-1,m_F} \nonumber\\
&= -\dfrac{\mu_B B}{2}(g_J - g_I)\left(\dfrac{[(J+I+1)^2 - F^2][F^2-(J-I)^2}{F}\right)^{1/2} \label{eq:offdiag}\\
& \hspace{4cm}\times \left(\dfrac{F^2-m_F^2}{F(2F+1)(2F-1)}\right)^{1/2} \nonumber 
\end{align}
where $B$ is the projection of the magnetic field $\vec B$ on the quantization axis. 
For the diagonal terms given by (\ref{eq:diag}), $E_0(F)$ is the zero-field energy of the $\ket{F,m_F}$ sub-level, $g_F$ is the associated Landé factor and $m_F$ is the magnetic quantum number. It is important to note that the Bohr magneton has been chosen with a negative sign \cite{RevModPhys.49.31} so that the model remains consistent. The off-diagonal terms are given by (\ref{eq:offdiag}), where $g_J$ and $g_I$ are respectively the total angular and nuclear Landé factors. These elements are non-zero only if they respect the selection rules $\Delta L =0$, $\Delta J = 0$, $\Delta F = \pm 1$ and $\Delta m_F = 0$. A consequence of the two previous formula is that the Hamiltonian $\mathcal{H}$ has a block diagonal structure where each block corresponds to a given value of $m_F$. This structure is similar to the one observed for the $D_1$ and $D_2$ lines of Rubidium \cite{hrant1}. In the case of $5\prescript{2}{}{S}_{1/2}\rightarrow 6\prescript{2}{}{P}_{1/2}$, blocks have a maximum size of $2\times 2$.

\subsection{Transitions between two Zeeman sublevels}
After diagonalization of the Hamiltonian, one obtains eigenvectors that can be expressed as linear combinations of the unperturbated atomic states vectors
\begin{equation}
\ket{\Psi(F_g,m_{F_g})} = \sum_{F_g'} c_{F_gF_g'}\ket{F_g',m_{F_g}} \label{eq:eiggr}
\end{equation}
\begin{equation}
\ket{\Psi(F_e,m_{F_e})} = \sum_{F_e'} c_{F_eF_e'}\ket{F_e',m_{F_e}}\, . \label{eq:eigex}
\end{equation}
The calculation of the transition intensity $A_{eg}$ between a "ground" Zeeman sublevel $\ket{\Psi(F_g,m_{F_g})}$  and an "excited" Zeeman sublevel $\ket{\Psi(F_e,m_{F_e})}$ is equivalent to the calculation of the squared transfer coefficients modified by the presence of the magnetic field $B$. These coefficients are given by the following formula:
\begin{align}
a[\ket{\Psi(F_e,m_{F_e})}&;\ket{\Psi(F_g,m_{F_g})};q] \nonumber\\ 
=& \sum_{F_e',F_g'} c_{F_eF_e'}a(F_e',m_{F_e};F_g',m_{F_g};q)c_{F_gF_g'}\, .\label{eq:transcoef}
\end{align}
Equation (\ref{eq:transcoef}) results in a sum of $n_g\times n_e$ terms ($n_g$ and $n_e$ being respectively the number of $F_g$ and $F_e$ levels) resulting in a combination of different magnetic quantum numbers. This combination can be cancelled as we will show later, which could be seen as quantum interferences.

The coefficients $a(F_e,m_{F_e};F_g,m_{F_g};q)$ are the unperturbated transfer coefficients given by
\begin{align}
&a(F_e,m_{F_e};F_g,m_{F_g};q) = (-1)^{1+I+J_e+F_e+F_g-m_{F_e}} \nonumber\\
\times & \sqrt{2J_e+1}\sqrt{2F_e+1}\sqrt{2F_g+1}
\begin{pmatrix}
F_e & 1 & F_g \\ 
-m_{F_e} & q & m_{F_g}
\end{pmatrix} 
\begin{Bmatrix}
F_e & 1 & F_g \\ 
J_g & I & J_e
\end{Bmatrix} 
\end{align}
where parenthesis represent a Wigner 3j-symbol and curly brackets represent a Wigner 6j-symbol. The index $q$ depends on the polarization of the laser: $q=0$ for $\pi$ polarization and $q=\pm 1$ for $\sigma^{\pm}$ polarization. In this article, we will denote the transitions as $\ket{F_g,m_{F_g}} \rightarrow \ket{F_e,m_{F_e}}$. Such transitions are allowed (resp. so-called forbidden) if they obey (resp. disobey) the selection rule $\Delta F = 0,\pm 1$.
It is important to note that the notation $\ket{F_g,m_{F_g}} \rightarrow \ket{F_e,m_{F_e}}$ is an ambiguous notation since the application of a magnetic field causes a mixing of states, thus $\ket{F,m_F}$ is not a good basis in the meaning of eigenbases in intermediate magnetic fields, and neither is $\ket{m_I,m_J}$. 

\section{$5\prescript{2}{}{S}_{1/2} \rightarrow 6\prescript{2}{}{P}_{1/2}$ transitions}
\subsection{$\pi$ transitions}
\begin{figure}[H]
\centering
\begin{tikzpicture} [scale=0.75, decoration={coil,aspect=0.4,segment length=3mm,amplitude=3mm}]

\draw(3,0) -- (4,0);
\draw(4.5,0) -- (5.5,0);
\draw(6,0) -- (7,0);
\draw(7.5,0) -- (8.5,0);
\draw(9,0) -- (10,0);

\draw(1.5,1) -- (2.5,1);
\draw(3,1) -- (4,1);
\draw(4.5,1) -- (5.5,1);
\draw(6,1) -- (7,1);
\draw(7.5,1) -- (8.5,1);
\draw(9,1) -- (10,1);
\draw(10.5,1) -- (11.5,1);

\draw(3,4) -- (4,4);
\draw(4.5,4) -- (5.5,4);
\draw(6,4) -- (7,4);
\draw(7.5,4) -- (8.5,4);
\draw(9,4) -- (10,4);

\draw(1.5,5) -- (2.5,5);
\draw(3,5) -- (4,5);
\draw(4.5,5) -- (5.5,5);
\draw(6,5) -- (7,5);
\draw(7.5,5) -- (8.5,5);
\draw(9,5) -- (10,5);
\draw(10.5,5) -- (11.5,5);

\node at (12.5,0) {$F_g=2$};
\node at (12.5,1) {$F_g=3$};
\node at (12.5,4) {$F_e=2$};
\node at (12.5,5) {$F_e=3$};

\draw[->](2,1) -- (2,5);

\draw[->](3.2,0) -- (3.2,4);
\draw[->](3.4,0) -- (3.4,5);
\draw[->](3.6,1) -- (3.6,4);
\draw[->](3.8,1) -- (3.8,5);

\draw[->](4.7,0) -- (4.7,4);
\draw[->](4.9,0) -- (4.9,5);
\draw[->](5.1,1) -- (5.1,4);
\draw[->](5.3,1) -- (5.3,5);

\draw[->](6.2,0) -- (6.2,4);
\draw[->](6.4,0) -- (6.4,5);
\draw[->](6.6,1) -- (6.6,4);
\draw[->](6.8,1) -- (6.8,5);

\draw[->](7.7,0) -- (7.7,4);
\draw[->](7.9,0) -- (7.9,5);
\draw[->](8.1,1) -- (8.1,4);
\draw[->](8.3,1) -- (8.3,5);

\draw[->](9.2,0) -- (9.2,4);
\draw[->](9.4,0) -- (9.4,5);
\draw[->](9.6,1) -- (9.6,4);
\draw[->](9.8,1) -- (9.8,5);

\draw[->](11,1) -- (11,5);

\node at (3.1,-0.2) {\tiny $1$};
\node at (3.5,-0.2) {\tiny $2$};
\node at (4.6,-0.2) {\tiny $3$};
\node at (5,-0.2) {\tiny $4$};
\node at (6.1,-0.2) {\tiny $5$};
\node at (6.5,-0.2) {\tiny $6$};
\node at (7.6,-0.2) {\tiny $7$};
\node at (8,-0.2) {\tiny $8$};
\node at (9.1,-0.2) {\tiny $9$};
\node at (9.5,-0.2) {\tiny $10$};

\node at (2,0.8) {\tiny $11$};
\node at (3.6,0.8) {\tiny $12$};
\node at (3.8,0.5) {\tiny $13$};
\node at (5.1,0.8) {\tiny $14$};
\node at (5.3,0.5) {\tiny $15$};
\node at (6.6,0.8) {\tiny $16$};
\node at (6.8,0.5) {\tiny $17$};
\node at (8.1,0.8) {\tiny $18$};
\node at (8.3,0.5) {\tiny $19$};
\node at (9.6,0.8) {\tiny $20$};
\node at (9.8,0.5) {\tiny $21$};
\node at (11,0.8) {\tiny $22$};

\node at (2,5.4) {\small $-3$};
\node at (3.5,5.4) {\small $-2$};
\node at (5,5.4) {\small $-1$};
\node at (6.5,5.4) {\small $0$};
\node at (8,5.4) {\small $+1$};
\node at (9.5,5.4) {\small $+2$};
\node at (11,5.4) {\small $+3$};

 \end{tikzpicture}
\caption{Possible $5\prescript{2}{}{S}_{1/2} \rightarrow 6\prescript{2}{}{P}_{1/2}$ $\pi$ transitions of ${}^{85}\text{Rb}\label{fig:schemepi85rb}$.}
\end{figure}
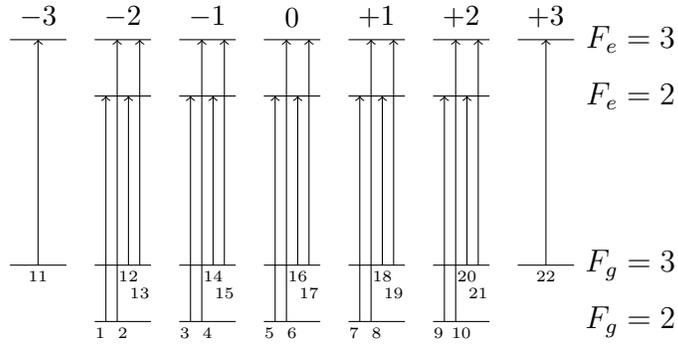

On figures \ref{fig:schemepi87rb} and \ref{fig:schemepi85rb} are schematized all the possible $5\prescript{2}{}{S}_{1/2} \rightarrow 6\prescript{2}{}{P}_{1/2}$ $\pi$ transitions of ${}^{87}\text{Rb}$ and ${}^{85}\text{Rb}$. In this case, there are no forbidden transitions. 

\begin{figure}[H]
\centering
\begin{tikzpicture} [scale=0.75, decoration={coil,aspect=0.4,segment length=3mm,amplitude=3mm}]


\draw(3,0) -- (4,0);
\draw(4.5,0) -- (5.5,0);
\draw(6,0) -- (7,0);

\draw(1.5,1) -- (2.5,1);
\draw(3,1) -- (4,1);
\draw(4.5,1) -- (5.5,1);
\draw(6,1) -- (7,1);
\draw(7.5,1) -- (8.5,1);

\draw(3,4) -- (4,4);
\draw(4.5,4) -- (5.5,4);
\draw(6,4) -- (7,4);

\draw(1.5,5) -- (2.5,5);
\draw(3,5) -- (4,5);
\draw(4.5,5) -- (5.5,5);
\draw(6,5) -- (7,5);
\draw(7.5,5) -- (8.5,5);

\node at (9.5,0) {$F_g=1$};
\node at (9.5,1) {$F_g=2$};
\node at (9.5,4) {$F_e=1$};
\node at (9.5,5) {$F_e=2$};

\node at (2,5.4) {\small $-2$};
\node at (3.5,5.4) {\small $-1$};
\node at (5,5.4) {\small $0$};
\node at (6.5,5.4) {\small $+1$};
\node at (8,5.4) {\small $+2$};

\draw[->](2,1) -- (2,5);

\draw[->](3.2,0) -- (3.2,4);
\draw[->](3.4,0) -- (3.4,5);
\draw[->](3.6,1) -- (3.6,4);
\draw[->](3.8,1) -- (3.8,5);

\draw[->](4.7,0) -- (4.7,4);
\draw[->](4.9,0) -- (4.9,5);
\draw[->](5.1,1) -- (5.1,4);
\draw[->](5.3,1) -- (5.3,5);

\draw[->](6.2,0) -- (6.2,4);
\draw[->](6.4,0) -- (6.4,5);
\draw[->](6.6,1) -- (6.6,4);
\draw[->](6.8,1) -- (6.8,5);

\draw[->](8,1) -- (8,5);

\node at (3.1,-0.2) {\tiny $1$};
\node at (3.5,-0.2) {\tiny $2$};
\node at (4.6,-0.2) {\tiny $3$};
\node at (5,-0.2) {\tiny $4$};
\node at (6.1,-0.2) {\tiny $5$};
\node at (6.5,-0.2) {\tiny $6$};

\node at (2,0.8) {\tiny $7$};
\node at (3.6,0.8) {\tiny $8$};
\node at (3.9,0.8) {\tiny $9$};
\node at (5.1,0.8) {\tiny $10$};
\node at (5.3,0.5) {\tiny $11$};
\node at (6.6,0.8) {\tiny $12$};
\node at (6.8,0.5) {\tiny $13$};
\node at (8,0.8) {\tiny $14$};

\end{tikzpicture}
\caption{Possible $5\prescript{2}{}{S}_{1/2} \rightarrow 6\prescript{2}{}{P}_{1/2}$ $\pi$ transitions of ${}^{87}\text{Rb}\label{fig:schemepi87rb}$.}
\end{figure}
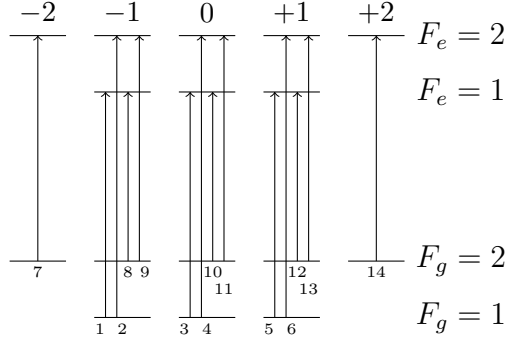

\begin{figure}[H]
\centering
\includegraphics[scale=0.6]{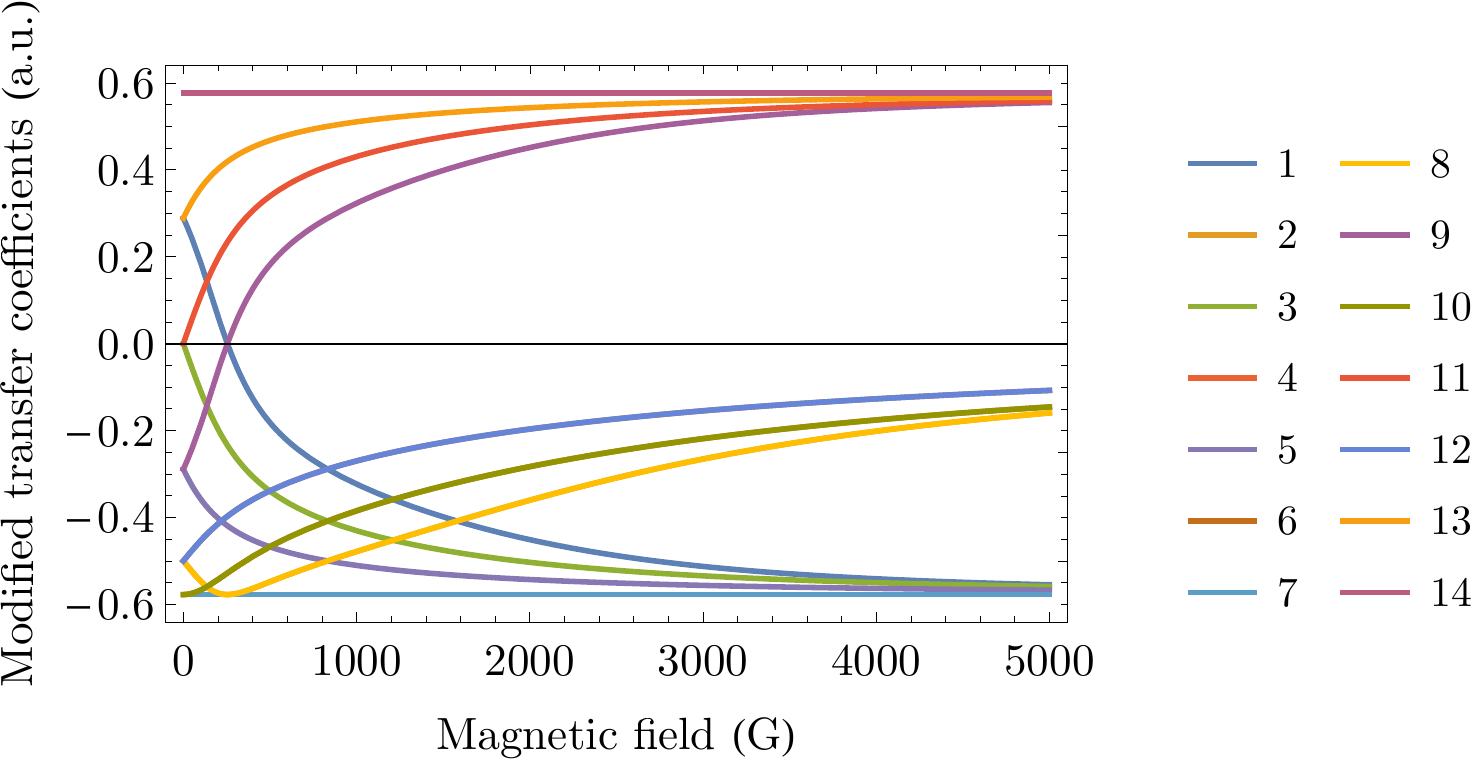}
\caption{$5\prescript{2}{}{S}_{1/2} \rightarrow 6\prescript{2}{}{P}_{1/2}$ $\pi$ transition modified transfer coefficients of ${}^{87}\text{Rb}$.\label{fig:transcoef_87rb_pi}}
\end{figure}

\begin{figure}[H]
\centering
\includegraphics[scale=0.6]{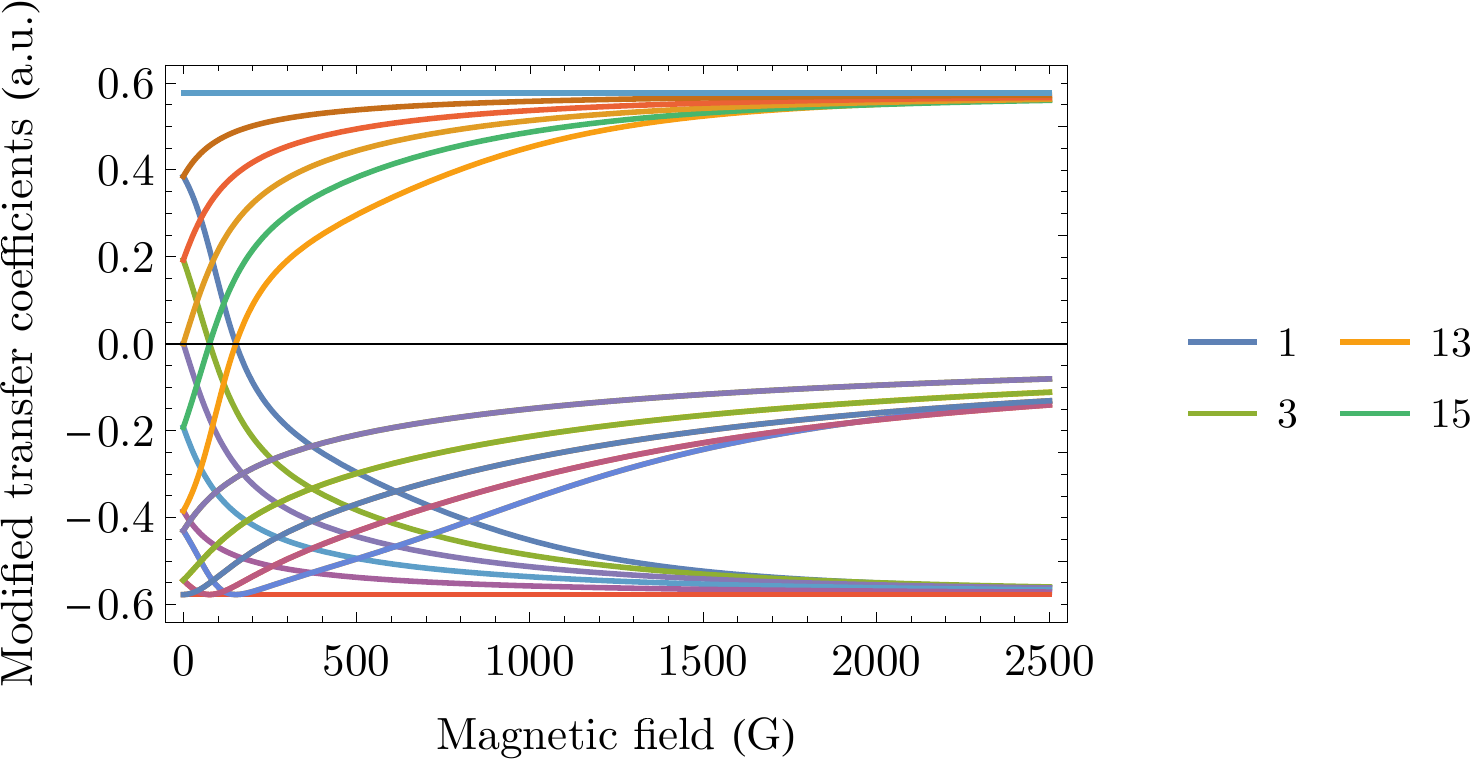}
\caption{$5\prescript{2}{}{S}_{1/2} \rightarrow 6\prescript{2}{}{P}_{1/2}$ $\pi$ transition modified transfer coefficients of ${}^{85}\text{Rb}$. For the sake of clarity, only the coefficients with a cancellation are labelled.\label{fig:transcoef_85rb_pi}}
\end{figure}

\noindent Using numerical simulations of the theoretical model described before, we can compute all the modified transfer coefficients $a[\ket{\Psi(F_e,m_{F_e})};\ket{\Psi(F_g,m_{F_g})};q]$ corresponding to these transitions. The results are shown on figures \ref{fig:transcoef_87rb_pi} and \ref{fig:transcoef_85rb_pi} for both isotopes. 
\noindent 
We will first focus on ${}^{87}\text{Rb}$. On figure \ref{fig:transcoef_87rb_pi} we notice that only two transfer coefficients cross the $x$-axis, meaning they are the only ones to cancel for a specific value of the external magnetic field $B$. These transfer coefficients correspond to the transitions labelled $1$ and $9$ (respectively $\ket{F_g=1,m_{F_g}=-1}\rightarrow \ket{F_e =1,m_{F_e} = -1}$ and $\ket{F_g=2,m_{F_g}=-1}\rightarrow \ket{F_e =2,m_{F_e} = -1}$). It is important to keep in mind that this notation is not the best since $\ket{F,m_F}$ is not a good basis in intermediate magnetic fields. The transfer coefficients have been computed using the following numerical data: $g_s~=~2.00231930436256(35)$ \cite{steck87}, $\mu_B~=~-1.39962449361(42)~\text{MHz/G}$ \cite{nist}, $g_I~=~-0.0009951414(10)$ and $g_L~=~0.99999369$ \cite{RevModPhys.49.31}, and the splittings displayed in table \ref{tab:hf_splittings}. Note that, on figure \ref{fig:transcoef_87rb_pi}, we can see constant lines corresponding to the transitions labelled 7 ($\ket{F_g = 2,m_{F_g} = -2} \rightarrow \ket{F_e = 2,m_{F_e} = -2}$) and 14 ($\ket{F_g = 2,m_{F_g} = +2} \rightarrow \ket{F_e = 2,m_{F_e} = +2}$). This is simply due to the fact that they correspond to transitions to states that remain pure (non-mixed) and do not depend on the value of $B$. 
Since we are interested in transition cancellations, we will now focus on the states corresponding to $m_{F_g} = m_{F_e} = -1$ and derive analytical formulas. 
We do not need to compute the total Hamiltonian, it is sufficient to compute the two $2 \times 2$ blocks corresponding to the ground and excited states that we study. Using (\ref{eq:diag}) and (\ref{eq:offdiag}), we obtain
\begin{equation} G = 
\begin{pmatrix}
\mu_B B (2g_I - g_g) & \sqrt{3}\mu_B B (g_I - g_g) \\ 
\sqrt{3}\mu_B B (g_I - g_g) & \mu_B g_g B + \zeta 
\end{pmatrix} 
\label{eq:groundham}
\end{equation}
\begin{equation} E = 
\begin{pmatrix}
\mu_B B (2g_I - g_e) & \sqrt{3}\mu_B B (g_I - g_e) \\ 
\sqrt{3}\mu_B B (g_I - g_e) & \mu_B g_e B + \zeta ' 
\end{pmatrix} 
\label{eq:exham}
\end{equation}
where we denoted $g_g = \frac{3g_I}{4} + \frac{g_s}{4}$ and $g_e = \frac{3g_I}{4} + \frac{g_L}{3}-\frac{g_s}{12}$. The matrices $G$ and $E$ are written in the basis $\ket{F,m_F}$. After diagonalization, we obtain the following eigenvalues:
\begin{equation*}
\lambda_{g\pm} = \dfrac{\zeta + 2\mu_B B g_I \pm [(\zeta + 2\mu_B B(g_g - g_I))^2+12\mu_B (g_g-g_I)B^2]^\frac{1}{2} }{2}
\end{equation*}
\begin{equation*}
\lambda_{e\pm} = \dfrac{\zeta ' + 2\mu_B B g_I \pm [(\zeta + 2\mu_B B(g_e - g_I))^2+12\mu_B (g_e-g_I)B^2]^\frac{1}{2} }{2}
\end{equation*}
where $\lambda_{g\pm}$ are the two eigenvalues of $G$ and $\lambda_{e\pm}$ are the two eigenvalues of $E$. The eigenvectors can be decomposed on the basis of the unperturbated atomic state vectors according to (\ref{eq:eiggr}) and (\ref{eq:eigex}). In this case, we obtain:
\begin{equation}
\ket{\Psi(F_g,m_{F_g})\pm} = \dfrac{\eta_{g\pm}}{\sqrt{1+\eta_{g\pm}^2}}\ket{1,-1} +  \dfrac{1}{\sqrt{1+\eta_{g\pm}^2}}\ket{2,-1}
\end{equation}
\begin{equation}
\ket{\Psi(F_e,m_{F_e})\pm} = \dfrac{\eta_{e\pm}}{\sqrt{1+\eta_{e\pm}^2}}\ket{1,-1} +  \dfrac{1}{\sqrt{1+\eta_{e\pm}^2}}\ket{2,-1}
\end{equation}
where we denoted 
\begin{equation*}
\eta_{g\pm} = \dfrac{\lambda_{g\pm} - \zeta - \mu_B g_g B}{\sqrt{3}\mu_B B(g_I - g_g)}
\end{equation*}
\begin{equation*}
\eta_{e\pm} = \dfrac{\lambda_{e\pm} - \zeta ' - \mu_B g_e B}{\sqrt{3}\mu_B B(g_I - g_e)}\, .
\end{equation*}
Making use of (\ref{eq:transcoef}), we can write the modified transfer coefficients as
\begin{align*}
&a[\ket{\Psi(F_e,m_{F_e})\pm};\ket{\Psi(F_g,m_{F_g})\pm};q=0] \nonumber \\
&=\left[\sqrt{3}(\eta_{g\pm}\eta_{e\pm} -1) + 3(\eta_{g\pm} +\eta_{e\pm})\right]\dfrac{1}{6\sqrt{1+\eta_{g\pm}^2}\sqrt{1+\eta_{e\pm}^2}}\, .
\end{align*}
These two transfer coefficients get cancelled for the same value of $B$ as shown by the numerical simulation. We can determine that the $B$-value causing the transfer coefficients to be cancelled are given by:
\begin{equation}
B^{(\pm)}_{(\pm)} = \dfrac{1}{\mu_B} \dfrac{3\zeta \zeta '}{3g_I\zeta -4g_L \zeta + g_s\zeta+3g_I\zeta ' - 3g_s \zeta '}\, .\label{eq:analytical}
\end{equation}
This formula is analogous to the one determined by Aleksanyan \textit{et al.} \cite{aleksanyan} for the $D_1$ line of ${}^{87}\text{Rb}$. In table \ref{tab:values87rb} we present the $B$-value obtained for both transitions using relation \eqref{eq:analytical}, where $B$ is the value with its uncertainty and $B^*$ is the value obtained when ignoring the uncertainty of the excited states' energy difference (ESED) $\zeta '$. We notice that the major loss of precision on $B$ comes from the small number of digits of $\zeta '$. In this case, the numerical simulation provides a value of $B = 254.423942950~\text{G}$ (all uncertainties ignored) which appears to be in perfect agreement with the formula we have derived (\ref{eq:analytical}). When recording the reflection or absorption spectra of an atomic vapor confined in a nano-cell, it is possible to measure the intensity of a transition and thus determine precisely the value of the external magnetic field when this transition cancels. This method would be even more efficient if an experiment was done to refine the value of $\zeta '$. Consequently, such values can be implemented as a standard for precise magnetometers.

\begin{table}[hbtp]
\centering
\begin{tabular}{|c|c|c|c|}
\hline

Atom                                 & Transitions & $B$(G)                       & $B^*$(G)                             \\ \hline
                                     & $1$         &                              &                                      \\ \cline{2-2}
\multirow{-2}{*}{${}^{87}\text{Rb}$} & $9$         & \multirow{-2}{*}{$254.42(39)$} & \multirow{-2}{*}{$254.423942950(79)$} \\ \hline
\end{tabular}
\caption{$B$-values cancelling ${}^{87}\text{Rb}$ $5\prescript{2}{}{S}_{1/2} \rightarrow 6\prescript{2}{}{P}_{1/2}$ $\pi$ transitions.}
\label{tab:values87rb}
\end{table} 

Analogously, on figure \ref{fig:transcoef_85rb_pi} we see that four transfer coefficients cross the $x$-axis. These coefficients correspond to the transitions labelled $1$, $3$, $13$ and $15$ (respectively $\ket{F_g = 2,m_{F_g} = -2}\rightarrow\ket{F_e = 2, m_{F_e} = -2}$, $\ket{F_g = 2,m_{F_g} = -1}\linebreak\rightarrow\ket{F_e = 2, m_{F_e} = -1}$, $\ket{F_g = 3,m_{F_g} = -2}\rightarrow\ket{F_e = 3, m_{F_e} = -2}$ and \linebreak $\ket{F_g = 3,m_{F_g} = -1}\rightarrow\ket{F_e = 3, m_{F_e} = -1}$). With the same type of calculations as before, focusing only on the blocks corresponding to $m_{F_g} = m_{F_e} = \pm 1$ and $m_{F_g} = m_{F_e} = \pm 2$, we obtain that the modified transfer coefficients mentioned before cancel for:

\begin{equation}
B_{(\pm)}^{(\pm)} = \dfrac{1}{\mu_B} \dfrac{2\xi \xi'}{3g_I\xi' - 3g_s \xi' + 3g_I \xi - 4g_L\xi + g_s\xi}~\text{for}~m = -1 \label{eq12}
\end{equation}
\begin{equation}
B_{(\pm)}^{(\pm)}  = \dfrac{1}{\mu_B} \dfrac{4\xi \xi'}{3g_I\xi' - 3g_s \xi' + 3g_I \xi - 4g_L\xi + g_s\xi}~\text{for}~m = -2\, .\label{eq13}
\end{equation}
The results obtained are shown in table \ref{tab:values85rb}

\begin{table}[hbtp]
\centering
\begin{tabular}{|c|c|c|c|}
\hline

Atom                                 & Transitions & $B$(G)                        & $B^*$(G)                            \\ \hline
                                     & $1$         &                               &                                     \\ \cline{2-2}
                                     & $13$        & \multirow{-2}{*}{$151.54(52)$}  & \multirow{-2}{*}{$151.54436391(17)$} \\ \cline{2-4} 
                                     & $3$         &                               &                                     \\ \cline{2-2}
\multirow{-4}{*}{${}^{85}\text{Rb}$} & $15$        & \multirow{-2}{*}{$75.77(26)$} & \multirow{-2}{*}{$75.772181955(86)$}   \\ \hline
\end{tabular}
\caption{$B$-values cancelling ${}^{85}\text{Rb}$ $5\prescript{2}{}{S}_{1/2} \rightarrow 6\prescript{2}{}{P}_{1/2}$ $\pi$ transitions}
\label{tab:values85rb}
\end{table}
As before, two different types of $B$-values are shown. $B$ is the value taking into account all the uncertainties and $B^*$ is the value obtained when ignoring the uncertainty on the ESED $\xi '$. Both $B$ and $B^*$ are obtained using \eqref{eq12} and \eqref{eq13}. Once again, we see that the numerical simulation and the theory are in perfect agreement (the values obtained numerically are $B = 75.772181955~\text{G}$ for $m=-1$ and $B = 151.54436391~\text{G}$ for $m=-2$). Consequently, we can use the numerical simulation to determine precisely the cancellations of transitions for which the calculations are heavier (when blocks are of size $3\times 3$ or higher, as it happens for $6\prescript{2}{}{P}_{3/2}$ states). Before looking at the case of $5\prescript{2}{}{S}_{1/2} \rightarrow 6\prescript{2}{}{P}_{3/2}$ transitions, we will just present graphs of the $\sigma^{\pm}$ transition intensities of both Rubidium isotopes.

\subsection{$\sigma^{\pm}$ transitions}
\begin{figure}[H]
\centering
\includegraphics[scale=0.65]{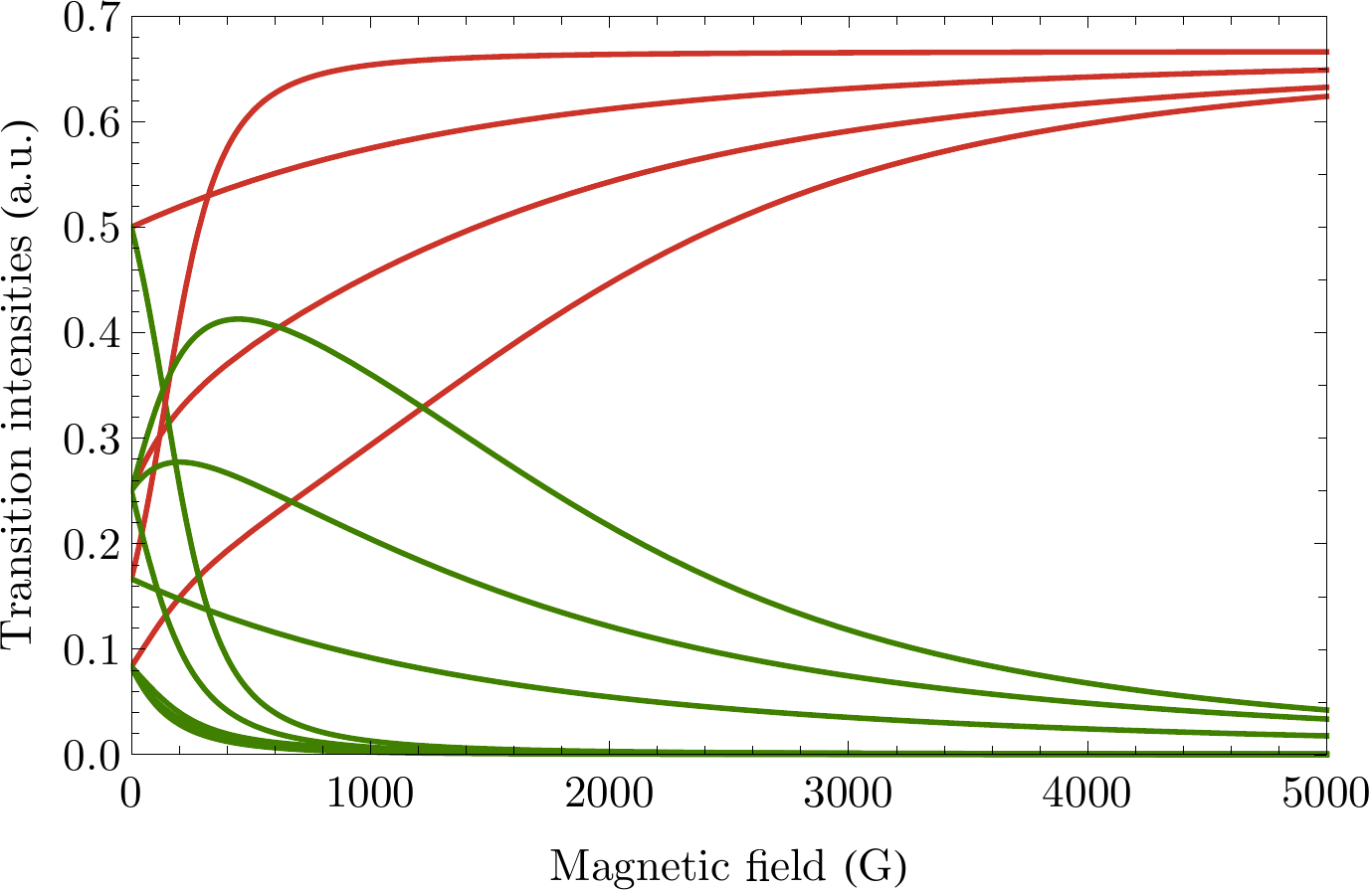} 
\includegraphics[scale=0.65]{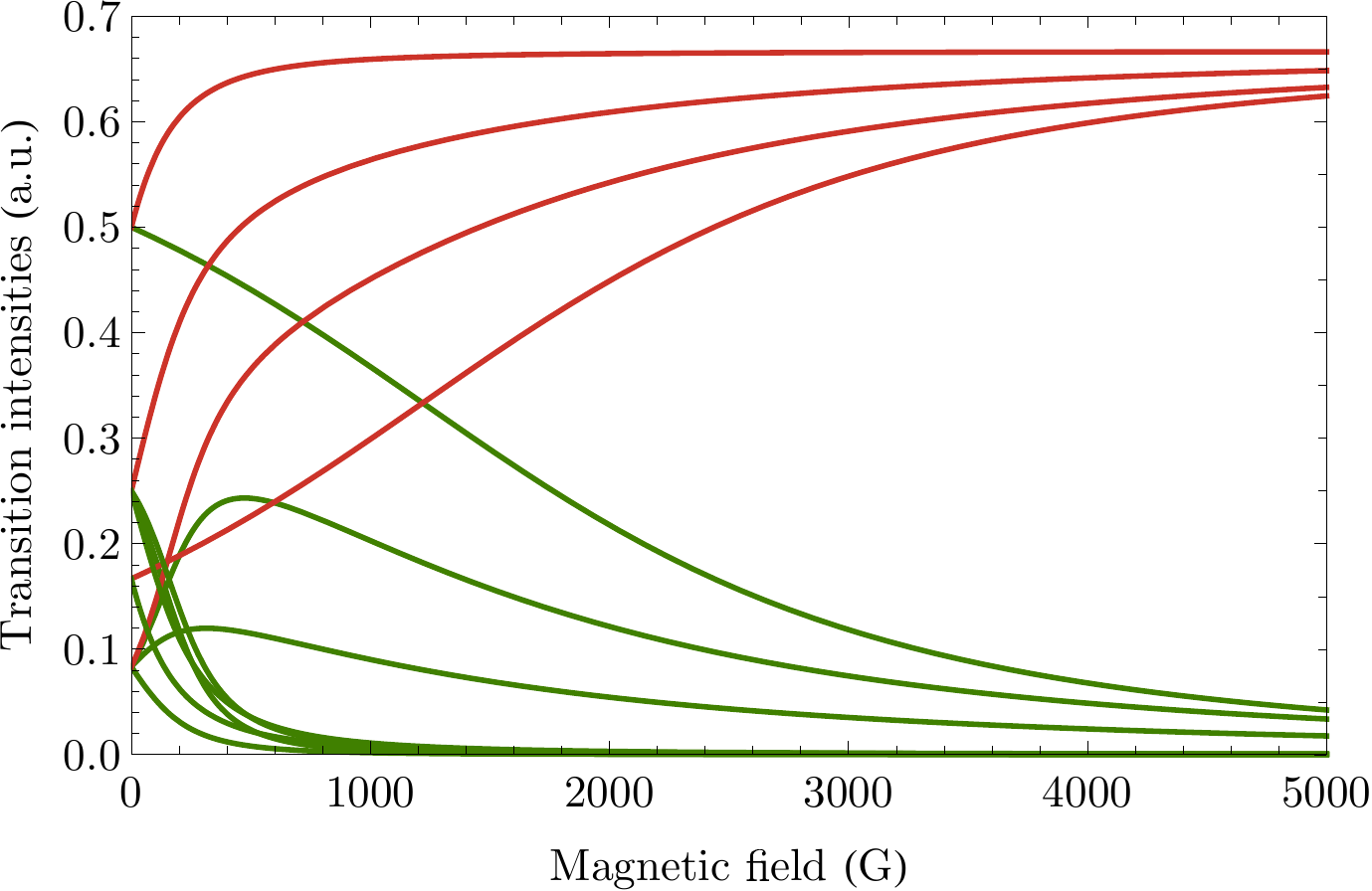}
\caption{${}^{87}\text{Rb}$ $5\prescript{2}{}{S}_{1/2} \rightarrow 6\prescript{2}{}{P}_{1/2}$ $\sigma^+$ (top) and $\sigma^-$ (bottom) transition intensities. \label{fig:87rbspm}}
\end{figure}

\begin{figure}[H]
\centering
\includegraphics[scale=0.65]{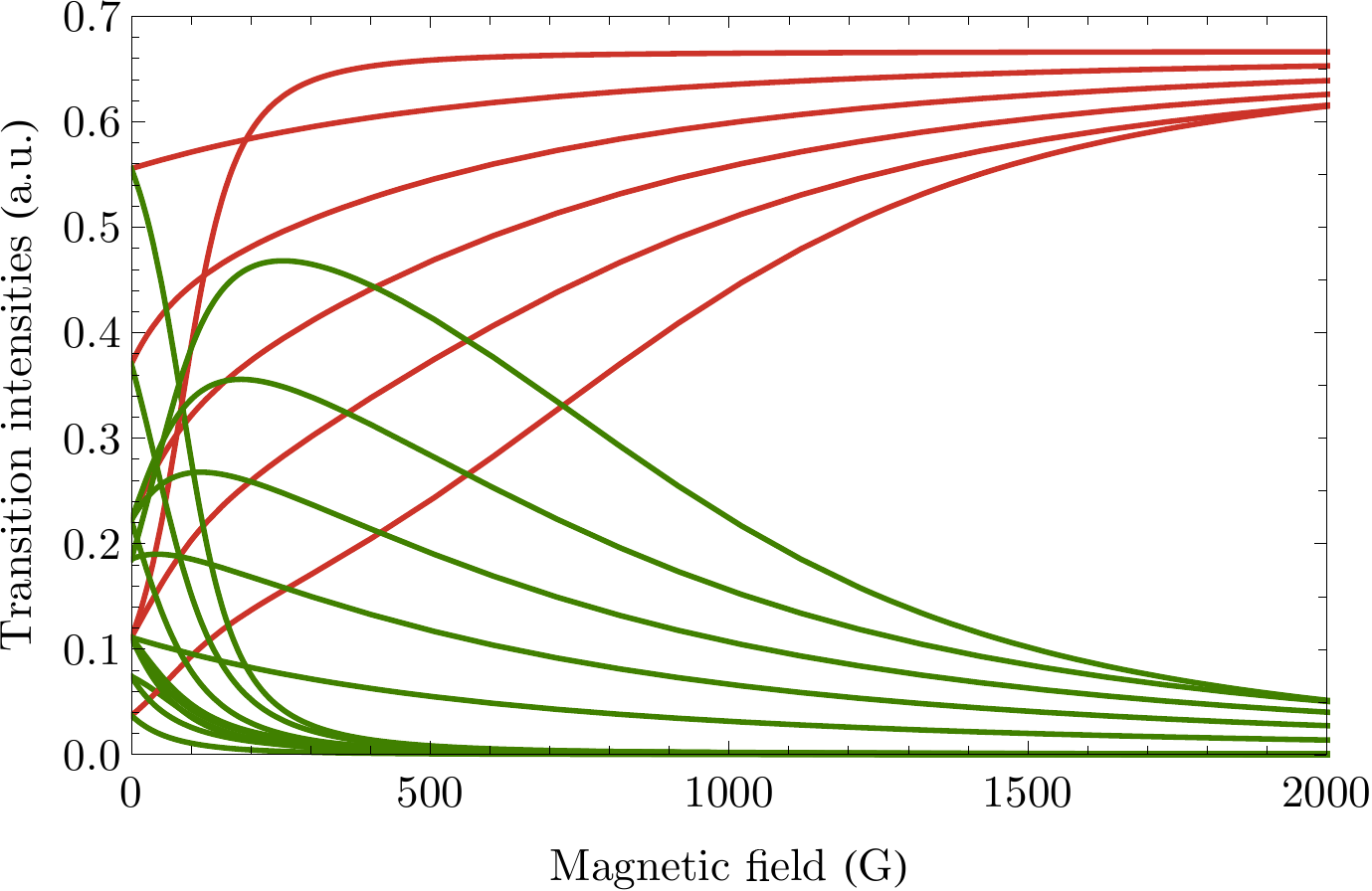}
\includegraphics[scale=0.65]{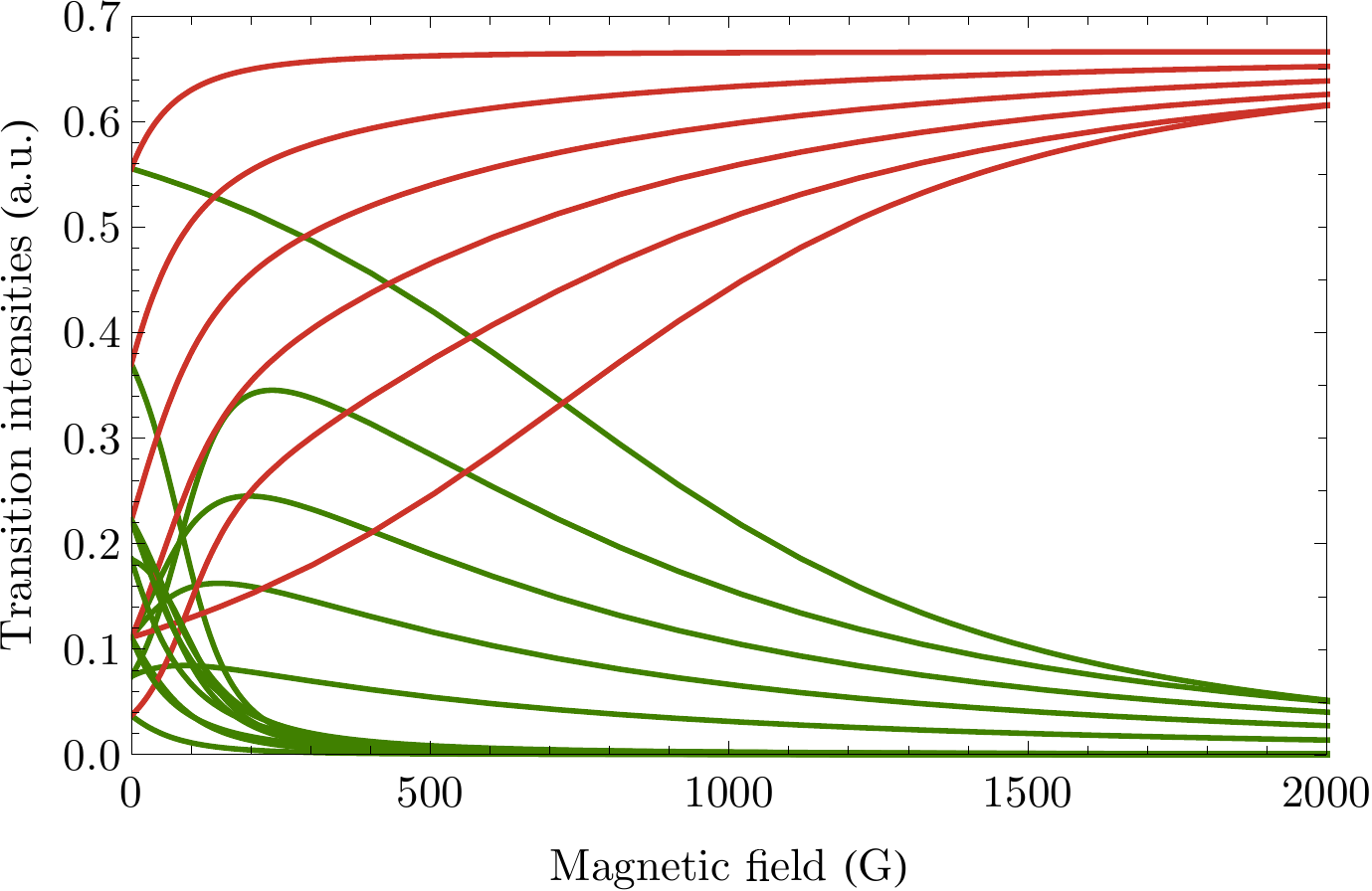}
\caption{${}^{85}\text{Rb}$ $5\prescript{2}{}{S}_{1/2} \rightarrow 6\prescript{2}{}{P}_{1/2}$ $\sigma^+$ (top) and $\sigma^-$ (bottom) transition intensities.\label{fig:85rbspm}}
\end{figure}

On figures \ref{fig:87rbspm} and \ref{fig:85rbspm} are shown all the $\sigma$ transition intensities between $5\prescript{2}{}{S}_{1/2}$ and $6\prescript{2}{}{P}_{1/2}$ states of ${}^{87}\text{Rb}$ and ${}^{85}\text{Rb}$.

Transitions that remain present in the spectra at high magnetic fields are represented in red and the ones that vanish are represented in green. For these transitions, we do not provide labelling since they are not interesting for our study. Indeed, none of these transition intensities are cancelled for specific values of $B$. 

To sum up, $38$ different transitions are possible for ${}^{87}\text{Rb}$ (resp. $62$ for ${}^{85}\text{Rb}$). Among all these transitions, only 2 (resp. 4) get cancelled for specific values of $B$, all these transitions corresponding to a $\pi$ polarization of the laser. Knowing that the numerical simulation is in full agreement with the theory, we will now use it to determine precisely cancellation of $5\prescript{2}{}{S}_{1/2} \rightarrow 6\prescript{2}{}{P}_{3/2}$ transitions for both isotopes.

\section{$5\prescript{2}{}{S}_{1/2} \rightarrow 6\prescript{2}{}{P}_{3/2}$ transitions of rubidium 87}
In this section, we will compute all the transition intensities of the possible $\pi$ and $\sigma^\pm$ transitions between $5\prescript{2}{}{S}_{1/2}$ and $6\prescript{2}{}{P}_{3/2}$. We will show graphs for the $\pi$ and $\sigma^\pm$ transitions of ${}^{87}\text{Rb}$. Due to the complexity of the calculations, we will not derive analytical formulas since cancelled transitions involve $3\times 3$ or $4\times 4$ blocks.

\subsection{$\pi$ transitions}

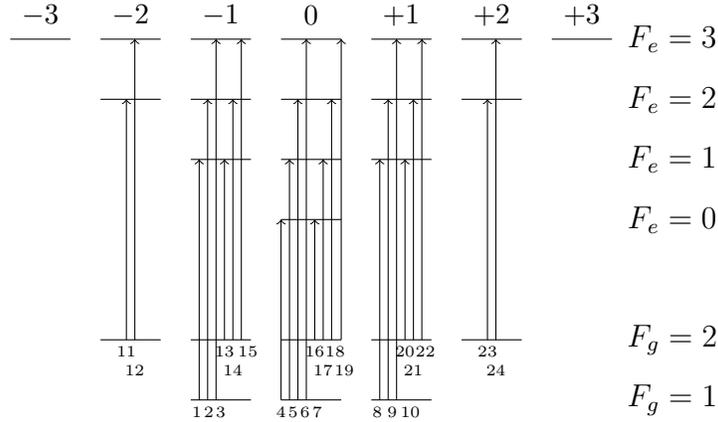
\begin{figure}[H]
\centering
\begin{tikzpicture} [scale=0.8, decoration={coil,aspect=0.4,segment length=3mm,amplitude=3mm}]


\draw(3,0) -- (4,0);
\draw(4.5,0) -- (5.5,0);
\draw(6,0) -- (7,0);

\draw(1.5,1) -- (2.5,1);
\draw(3,1) -- (4,1);
\draw(4.5,1) -- (5.5,1);
\draw(6,1) -- (7,1);
\draw(7.5,1) -- (8.5,1);

\draw(4.5,3) -- (5.5,3);

\draw(3,4) -- (4,4);
\draw(4.5,4) -- (5.5,4);
\draw(6,4) -- (7,4);

\draw(1.5,5) -- (2.5,5);
\draw(3,5) -- (4,5);
\draw(4.5,5) -- (5.5,5);
\draw(6,5) -- (7,5);
\draw(7.5,5) -- (8.5,5);

\draw(0,6) -- (1,6);
\draw(1.5,6) -- (2.5,6);
\draw(3,6) -- (4,6);
\draw(4.5,6) -- (5.5,6);
\draw(6,6) -- (7,6);
\draw(7.5,6) -- (8.5,6);
\draw(9,6) -- (10,6);

\node at (11,0) {$F_g=1$};
\node at (11,1) {$F_g=2$};
\node at (11,3) {$F_e=0$};
\node at (11,4) {$F_e=1$};
\node at (11,5) {$F_e=2$};
\node at (11,6) {$F_e=3$};

\node at (0.5,6.4) {\small $-3$};
\node at (2,6.4) {\small $-2$};
\node at (3.5,6.4) {\small $-1$};
\node at (5,6.4) {\small $0$};
\node at (6.5,6.4) {\small $+1$};
\node at (8,6.4) {\small $+2$};
\node at (9.5,6.4) {\small $+3$};

\draw[->](1.93,1) -- (1.93,5);
\draw[->](2.07,1) -- (2.07,6);

\draw[->](3.14,0) -- (3.14,4);
\draw[->](3.28,0) -- (3.28,5);
\draw[->](3.42,0) -- (3.42,6);
\draw[->](3.56,1) -- (3.56,4);
\draw[->](3.70,1) -- (3.70,5);
\draw[->](3.84,1) -- (3.84,6);

\draw[->](4.5,0) -- (4.5,3);
\draw[->](4.64,0) -- (4.64,4);
\draw[->](4.78,0) -- (4.78,5);
\draw[->](4.92,0) -- (4.92,6);
\draw[->](5.06,1) -- (5.06,3);
\draw[->](5.20,1) -- (5.20,4);
\draw[->](5.34,1) -- (5.34,5);
\draw[->](5.5,1) -- (5.5,6);

\draw[->](6.14,0) -- (6.14,4);
\draw[->](6.28,0) -- (6.28,5);
\draw[->](6.42,0) -- (6.42,6);
\draw[->](6.56,1) -- (6.56,4);
\draw[->](6.70,1) -- (6.70,5);
\draw[->](6.84,1) -- (6.84,6);

\draw[->](7.93,1) -- (7.93,5);
\draw[->](8.07,1) -- (8.07,6);

\node at (3.1,-0.2) {\tiny $1$};
\node at (3.3,-0.2) {\tiny $2$};
\node at (3.5,-0.2) {\tiny $3$};

\node at (4.5,-0.2) {\tiny $4$};
\node at (4.7,-0.2) {\tiny $5$};
\node at (4.9,-0.2) {\tiny $6$};
\node at (5.1,-0.2) {\tiny $7$};

\node at (6.1,-0.2) {\tiny $8$};
\node at (6.35,-0.2) {\tiny $9$};
\node at (6.65,-0.2) {\tiny $10$};

\node at (1.93,0.8) {\tiny $11$};
\node at (2.07,0.5) {\tiny $12$};

\node at (3.56,0.8) {\tiny $13$};
\node at (3.70,0.5) {\tiny $14$};
\node at (3.95,0.8) {\tiny $15$};

\node at (5.06,0.8) {\tiny $16$};
\node at (5.20,0.5) {\tiny $17$};
\node at (5.4,0.8) {\tiny $18$};
\node at (5.55,0.5) {\tiny $19$};

\node at (6.56,0.8) {\tiny $20$};
\node at (6.70,0.5) {\tiny $21$};
\node at (6.9,0.8) {\tiny $22$};

\node at (7.93,0.8) {\tiny $23$};
\node at (8.07,0.5) {\tiny $24$};

\end{tikzpicture}
\caption{Possible $5\prescript{2}{}{S}_{1/2} \rightarrow 6\prescript{2}{}{P}_{3/2}$ $\pi$ transitions of ${}^{87}\text{Rb}$.\label{fig:schemepi87rb_6p32}}
\end{figure}

\begin{figure}[H]
\centering
\includegraphics[scale=0.6]{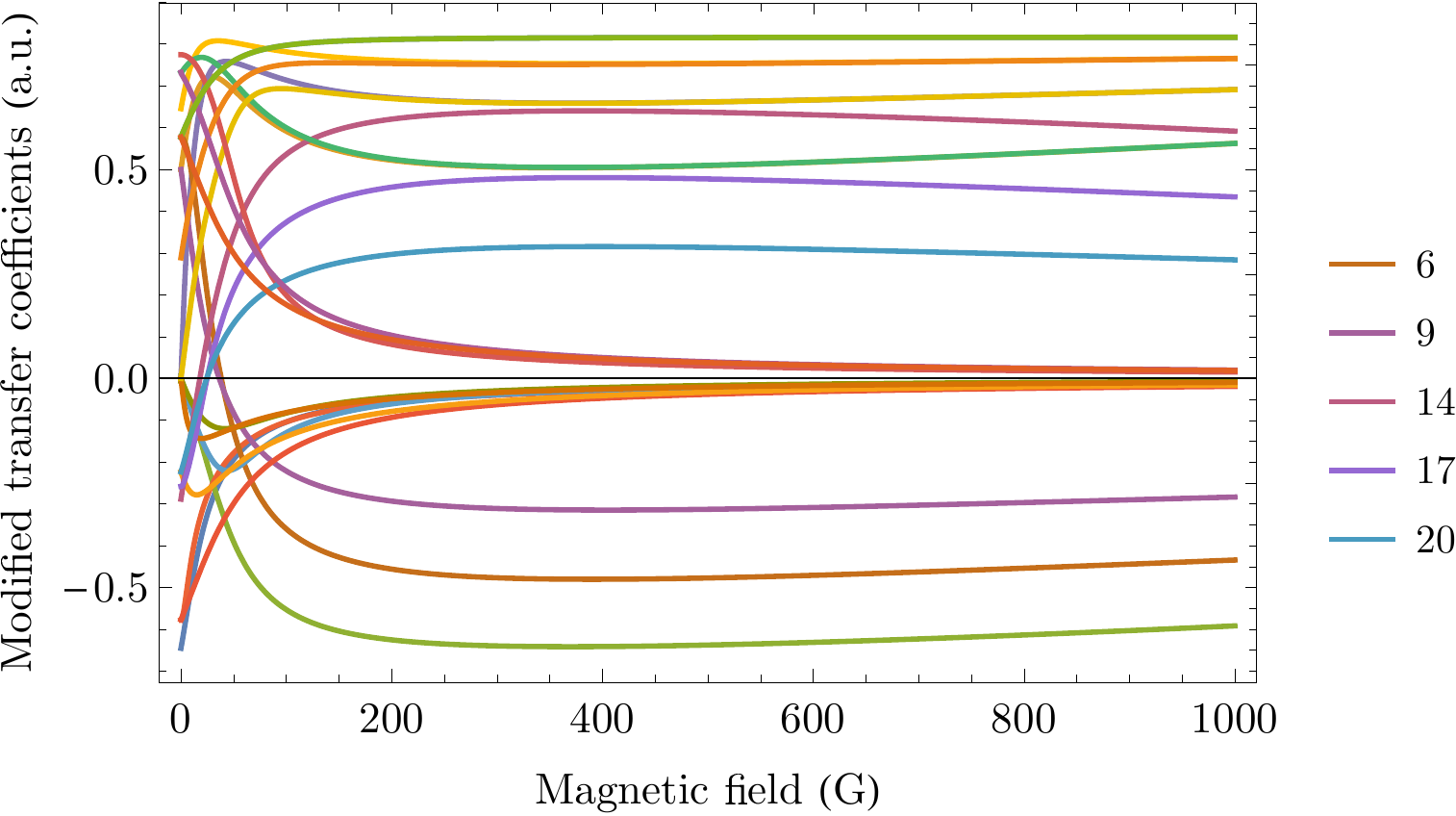}
\caption{$5\prescript{2}{}{S}_{1/2} \rightarrow 6\prescript{2}{}{P}_{3/2}$ $\pi$ transition modified transfer coefficients. For the sake of clarity, only the ones with a cancellation are labelled. \label{fig:transcoef_87rb_piP32}}
\end{figure}

All the modified transfer coefficients corresponding to the transitions labelled on figure \ref{fig:schemepi87rb_6p32} are represented on figure \ref{fig:transcoef_87rb_piP32}.
In this case, transitions $3$, $7$, $10$ and $16$ are forbidden in the general case but none of them cross the $x$-axis.

On this figure, all the curves vary according to the magnetic field since none of them correspond to transitions between two pure states. 
Among the 24 possible transitions, the transitions labelled $6$, $9$, $14$, $17$ and $20$ have a cancellation. These transitions have a magnetic quantum number of either $-1$, $0$ or $1$, unlike for $5\prescript{2}{}{S}_{1/2} \rightarrow 6\prescript{2}{}{P}_{1/2}$ transitions where we only had $m=-1$ for ${}^{87}\text{Rb}$. Here, each transition is cancelled for a different value of $B$. However, experimental measurements could be more difficult in this case due to the proximity of certain values. 

Table \ref{tab:values87rbpi32} shows the $B$-values for each cancellation, determined numerically with (3rd column) and without (4th column) taking into account the uncertainty on the ESEDs.

\begin{table}[hbtp]
\centering
\begin{tabular}{|c|c|c|c|}
\hline

Atom                                 & Transitions & $B$(G)        & $B^*$(G)           \\ \hline
                                     & $6$         & $38.21(19)$  & $38.211312878(40)$ \\ \cline{2-4} 
                                     & $9$         & $36.31(16)$  & $36.318455634(38)$ \\ \cline{2-4} 
                                     & $14$        & $17.895(66)$ & $17.895382415(19)$ \\ \cline{2-4} 
                                     & $17$        & $24.77(13)$  & $24.771185393(26)$ \\ \cline{2-4} 
\multirow{-5}{*}{${}^{87}\text{Rb}$} & $20$        & $24.36(11)$      & $24.361280734(26)$ \\ \hline
\end{tabular}
\caption{$B$-values cancelling ${}^{87}\text{Rb}$ $5\prescript{2}{}{S}_{1/2} \rightarrow 6\prescript{2}{}{P}_{3/2}$ $\pi$ transitions.}
\label{tab:values87rbpi32}
\end{table}

\subsection{$\sigma^+$ transitions}

\begin{figure}[H]
\centering
 \begin{tikzpicture} [scale=0.7, decoration={coil,aspect=0.4,segment length=3mm,amplitude=3mm}]


\draw(1.5,1) -- (2.5,1);
\draw(3,1) -- (4,1);
\draw(4.5,1) -- (5.5,1);
\draw(6,1) -- (7,1);
\draw(7.5,1) -- (8.5,1);

\draw(4.5,3) -- (5.5,3);

\draw(3,4) -- (4,4);
\draw(4.5,4) -- (5.5,4);
\draw(6,4) -- (7,4);

\draw(1.5,5) -- (2.5,5);
\draw(3,5) -- (4,5);
\draw(4.5,5) -- (5.5,5);
\draw(6,5) -- (7,5);
\draw(7.5,5) -- (8.5,5);

\draw(0,6) -- (1,6);
\draw(1.5,6) -- (2.5,6);
\draw(3,6) -- (4,6);
\draw(4.5,6) -- (5.5,6);
\draw(6,6) -- (7,6);
\draw(7.5,6) -- (8.5,6);
\draw(9,6) -- (10,6);

\node at (11,1) {$F_g=2$};
\node at (11,3) {$F_e=0$};
\node at (11,4) {$F_e=1$};
\node at (11,5) {$F_e=2$};
\node at (11,6) {$F_e=3$};

\node at (0.5,6.4) {\small $-3$};
\node at (2,6.4) {\small $-2$};
\node at (3.5,6.4) {\small $-1$};
\node at (5,6.4) {\small $0$};
\node at (6.5,6.4) {\small $+1$};
\node at (8,6.4) {\small $+2$};
\node at (9.5,6.4) {\small $+3$};

\draw[->](3.2-1.5,1) -- (5.30-1.5,6);
\draw[->](3.40-1.5,1) -- (5.10-1.5,5);
\draw[->](3.60-1.5,1) -- (4.90-1.5,4);

\draw[->](3.2,1) -- (5.30,6);
\draw[->](3.40,1) -- (5.10,5);
\draw[->](3.60,1) -- (4.90,4);
\draw[->](3.80,1) -- (4.70,3);

\draw[->](3.2+1.5,1) -- (5.30+1.5,6);
\draw[->](3.40+1.5,1) -- (5.10+1.5,5);
\draw[->](3.60+1.5,1) -- (4.90+1.5,4);

\draw[->](3.2+1.5+1.5,1) -- (5.30+1.5+1.5,6);
\draw[->](3.40+1.5+1.5,1) -- (5.10+1.5+1.5,5);

\draw[->](3.2+1.5+1.5+1.5,1) -- (5.30+1.5+1.5+1.5,6);

\node at (3.2+1.5+1.5+1.5,0.8) {\tiny $22$};

\node at (3.2+1.5+1.5,0.8) {\tiny $21$};
\node at (3.4+1.5+1.5,0.5) {\tiny $20$};

\node at (3.2+1.5,0.8) {\tiny $19$};
\node at (3.4+1.5,0.5) {\tiny $18$};
\node at (3.6+1.5,0.8) {\tiny $17$};

\node at (3.2,0.8) {\tiny $16$};
\node at (3.4,0.5) {\tiny $15$};
\node at (3.6,0.8) {\tiny $14$};
\node at (3.8,0.5) {\tiny $13$};

\node at (3.2-1.5,0.8) {\tiny $12$};
\node at (3.4-1.5,0.5) {\tiny $11$};
\node at (3.6-1.5,0.8) {\tiny $10$};

\end{tikzpicture}
\caption{Possible $5\prescript{2}{}{S}_{1/2} \rightarrow 6\prescript{2}{}{P}_{3/2}$ $\sigma^+$ transitions of ${}^{87}\text{Rb}$, $F_g = 2$.\label{fig:2schemesp87rb_6p32}}
\end{figure}
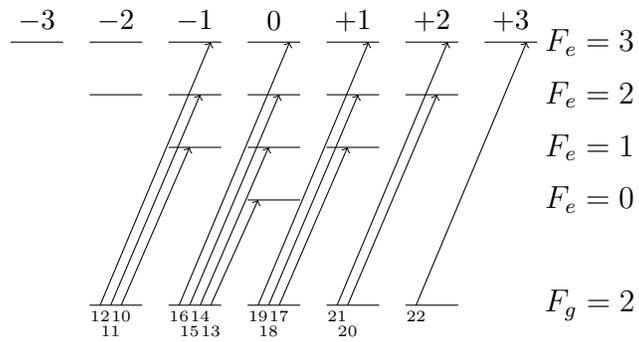

\begin{figure}[H]
\centering
\includegraphics[scale=0.6]{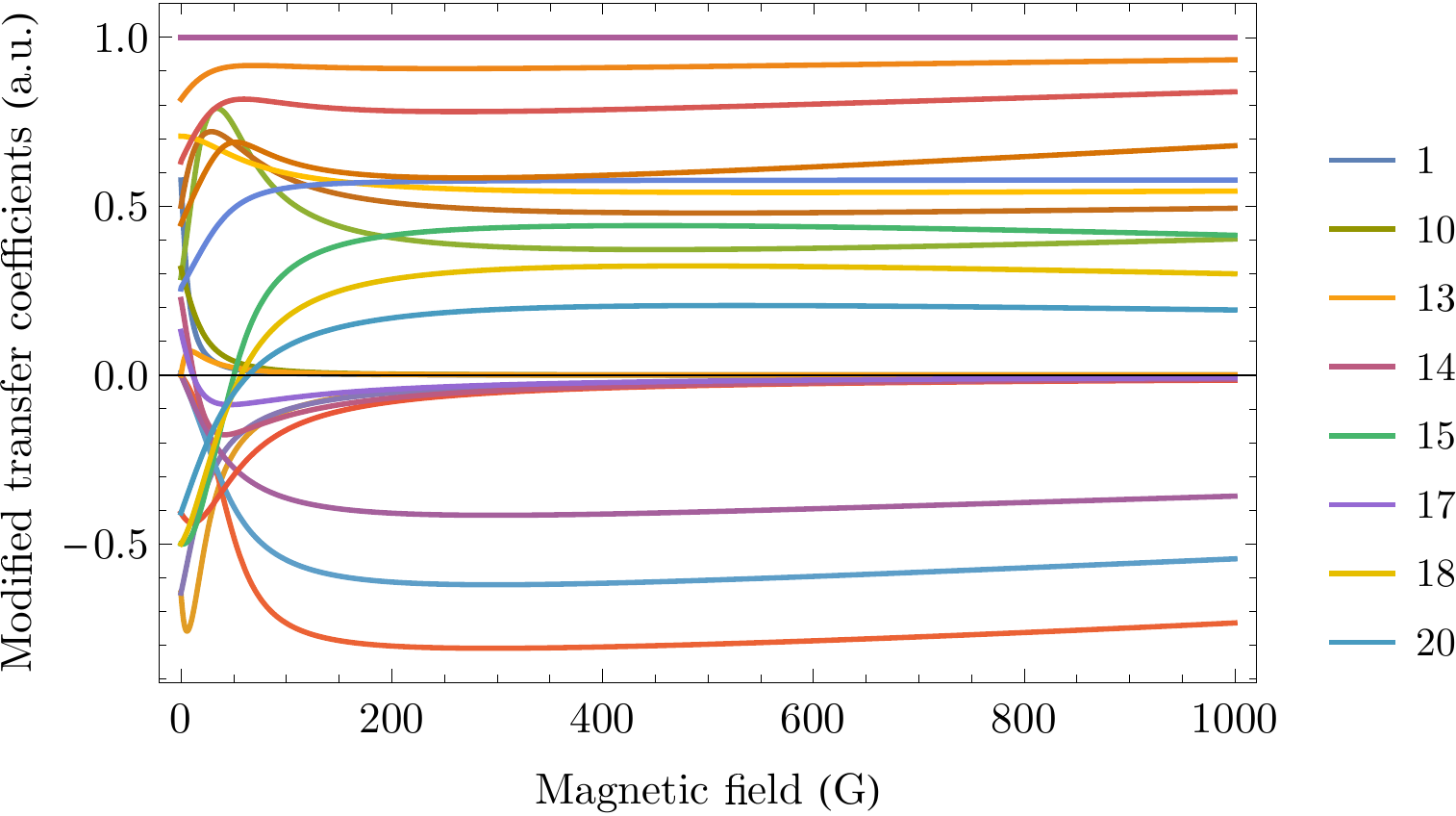}
\caption{$5\prescript{2}{}{S}_{1/2}\rightarrow 6\prescript{2}{}{P}_{3/2}$ $\sigma^+$ modified transfer coefficients. For the sake of clarity, only the ones with a cancellation are labelled. \label{fig:transcoef_87rb_spP32}}
\end{figure}

All the modified transfer coefficients corresponding to the transitions labelled on figures \ref{fig:schemesp87rb_6p32} and \ref{fig:2schemesp87rb_6p32}  are represented on figure \ref{fig:transcoef_87rb_spP32}.

\begin{figure}[H]
\centering
 \begin{tikzpicture} [scale=0.7, decoration={coil,aspect=0.4,segment length=3mm,amplitude=3mm}]


\draw(3,1) -- (4,1);
\draw(4.5,1) -- (5.5,1);
\draw(6,1) -- (7,1);

\draw(4.5,3) -- (5.5,3);

\draw(3,4) -- (4,4);
\draw(4.5,4) -- (5.5,4);
\draw(6,4) -- (7,4);

\draw(1.5,5) -- (2.5,5);
\draw(3,5) -- (4,5);
\draw(4.5,5) -- (5.5,5);
\draw(6,5) -- (7,5);
\draw(7.5,5) -- (8.5,5);

\draw(0,6) -- (1,6);
\draw(1.5,6) -- (2.5,6);
\draw(3,6) -- (4,6);
\draw(4.5,6) -- (5.5,6);
\draw(6,6) -- (7,6);
\draw(7.5,6) -- (8.5,6);
\draw(9,6) -- (10,6);

\node at (11,1) {$F_g=1$};
\node at (11,3) {$F_e=0$};
\node at (11,4) {$F_e=1$};
\node at (11,5) {$F_e=2$};
\node at (11,6) {$F_e=3$};

\node at (0.5,6.4) {\small $-3$};
\node at (2,6.4) {\small $-2$};
\node at (3.5,6.4) {\small $-1$};
\node at (5,6.4) {\small $0$};
\node at (6.5,6.4) {\small $+1$};
\node at (8,6.4) {\small $+2$};
\node at (9.5,6.4) {\small $+3$};

\draw[->](3.2,1) -- (5.30,6);
\draw[->](3.40,1) -- (5.10,5);
\draw[->](3.60,1) -- (4.90,4);
\draw[->](3.80,1) -- (4.70,3);

\draw[->](3.2+1.5,1) -- (5.30+1.5,6);
\draw[->](3.40+1.5,1) -- (5.10+1.5,5);
\draw[->](3.60+1.5,1) -- (4.90+1.5,4);

\draw[->](3.2+1.5+1.5,1) -- (5.30+1.5+1.5,6);
\draw[->](3.40+1.5+1.5,1) -- (5.10+1.5+1.5,5);

\node at (3.2,0.8) {\tiny $4$};
\node at (3.4,0.8) {\tiny $3$};
\node at (3.6,0.8) {\tiny $2$};
\node at (3.8,0.8) {\tiny $1$};

\node at (3.2+1.5,0.8) {\tiny $7$};
\node at (3.4+1.5,0.8) {\tiny $6$};
\node at (3.6+1.5,0.8) {\tiny $5$};

\node at (3.2+1.5+1.5,0.8) {\tiny $9$};
\node at (3.4+1.5+1.5,0.8) {\tiny $8$};

\end{tikzpicture}
\caption{Possible $5\prescript{2}{}{S}_{1/2} \rightarrow 6\prescript{2}{}{P}_{3/2}$ $\sigma^+$ transitions of ${}^{87}\text{Rb}$, $F_g = 1$.\label{fig:schemesp87rb_6p32}}
\end{figure}
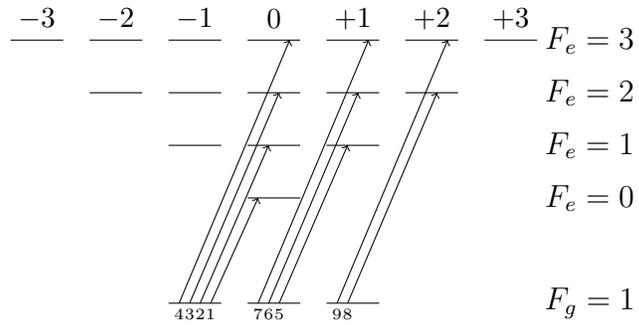

\noindent
Eight $\sigma^+$ transitions (respectively labelled $1$, $10$, $13$, $14$, $15$, $17$, $18$ and $20$, according to figures \ref{fig:schemesp87rb_6p32} and \ref{fig:2schemesp87rb_6p32}) are cancelled for a certain value of $B$. Transition $13$ is a forbidden transition. 
The $B$-values are presented in table \ref{tab:values87rbsp32}. Note the important number of cancellations compared to  
 the case of $5\prescript{2}{}{S}_{1/2} \rightarrow 6\prescript{2}{}{P}_{1/2}$ where no cancellation could be observed for $\sigma^\pm$. We notice that the three $B$-values of transitions $1$, $10$ and $13$ have much bigger uncertainties than the others.

For transition 20, we were able to exhibit the following analytical formula (among all these transitions it is the only one where Hamiltonians are of maximum $2\times 2$ dimension):

\begin{equation}
B = -\frac{1}{\mu_B}\frac{4 \gamma  \zeta  (3 g_I (\gamma -\zeta )+2 \zeta  g_L+g_S (\zeta -3 \gamma ))}{c_1 c_2}
\end{equation}
where we denoted $c_1 = 6 \gamma  g_I-3 \zeta  g_I+2 \zeta  g_L-6 \gamma  g_S+\zeta g_S$ and $c_2 = 2 \gamma  g_I-3 \zeta  g_I+2 \zeta  g_L-2 \gamma g_S+\zeta  g_S$. This formula provides a $B$-value for the cancellation $B = 64.133588295~\text{G}$ (all uncertainties ignored) showing the theory to be in perfect agreement with the simulation. 

\begin{table}[H]
\centering
\begin{tabular}{|c|c|c|c|}
\hline

Atom                                 & Transitions & $B$(G)                                       & \multicolumn{1}{c|}{$B^*$(G)} \\ \hline
                                     & $1$         & \textbf{$524(68)$} & $517.98168965(55)$                                    \\ \cline{2-4} 
                                     & $10$        & \textbf{$633(45)$} & $630.07368077(67)$                                    \\ \cline{2-4} 
                                     & $13$        &\textbf{$606(45)$} & $603.37498565(64)$                                    \\ \cline{2-4} 
                                     & $14$        & $12.219(77)$                                & $12.219830989(13)$                                    \\ \cline{2-4} 
                                     & $15$        & $50.30(20)$                                 & $50.306012284(52)$                                    \\ \cline{2-4} 
                                     & $17$        & $11.259(66)$                                & $11.259065240(12)$                                    \\ \cline{2-4} 
                                     & $18$        & $57.11(20)$                                 & $57.111350606(60)$                                    \\ \cline{2-4} 
\multirow{-8}{*}{${}^{87}\text{Rb}$} & $20$        & $64.13(20)$                                 & $64.133588295(68)$                                    \\ \hline
\end{tabular}
\caption{$B$-values cancelling ${}^{87}\text{Rb}$ $5\prescript{2}{}{S}_{1/2} \rightarrow 6\prescript{2}{}{P}_{3/2}$ $\sigma^+$ transitions.}
\label{tab:values87rbsp32}
\end{table}

\subsection{$\sigma^-$ transitions}
 
 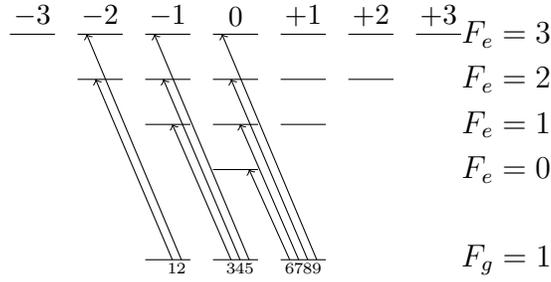
\begin{figure}[hbtp]
\centering
 \begin{tikzpicture} [scale=0.6, decoration={coil,aspect=0.4,segment length=3mm,amplitude=3mm}]


\draw(3,1) -- (4,1);
\draw(4.5,1) -- (5.5,1);
\draw(6,1) -- (7,1);

\draw(4.5,3) -- (5.5,3);

\draw(3,4) -- (4,4);
\draw(4.5,4) -- (5.5,4);
\draw(6,4) -- (7,4);

\draw(1.5,5) -- (2.5,5);
\draw(3,5) -- (4,5);
\draw(4.5,5) -- (5.5,5);
\draw(6,5) -- (7,5);
\draw(7.5,5) -- (8.5,5);

\draw(0,6) -- (1,6);
\draw(1.5,6) -- (2.5,6);
\draw(3,6) -- (4,6);
\draw(4.5,6) -- (5.5,6);
\draw(6,6) -- (7,6);
\draw(7.5,6) -- (8.5,6);
\draw(9,6) -- (10,6);

\node at (11,1) {$F_g=1$};
\node at (11,3) {$F_e=0$};
\node at (11,4) {$F_e=1$};
\node at (11,5) {$F_e=2$};
\node at (11,6) {$F_e=3$};

\node at (0.5,6.4) {\small $-3$};
\node at (2,6.4) {\small $-2$};
\node at (3.5,6.4) {\small $-1$};
\node at (5,6.4) {\small $0$};
\node at (6.5,6.4) {\small $+1$};
\node at (8,6.4) {\small $+2$};
\node at (9.5,6.4) {\small $+3$};

\draw[->](3.8,1) -- (4.70-3,6);
\draw[->](3.6,1) -- (4.90-3,5);

\draw[->](3.8+1.5,1) -- (4.70-1.5,6);
\draw[->](3.6+1.5,1) -- (4.90-1.5,5);
\draw[->](3.4+1.5,1) -- (5.10-1.5,4);

\draw[->](3.8+3,1) -- (4.70,6);
\draw[->](3.6+3,1) -- (4.90,5);
\draw[->](3.4+3,1) -- (5.10,4);
\draw[->](3.2+3,1) -- (5.30,3);

\node at (3.6,0.8) {\tiny$1$};
\node at (3.8,0.8) {\tiny$2$};

\node at (3.4+1.5,0.8) {\tiny$3$};
\node at (3.6+1.5,0.8) {\tiny$4$};
\node at (3.8+1.5,0.8) {\tiny$5$};

\node at (3.2+1.5+1.5,0.8) {\tiny$6$};
\node at (3.4+1.5+1.5,0.8) {\tiny$7$};
\node at (3.6+1.5+1.5,0.8) {\tiny$8$};
\node at (3.8+1.5+1.5,0.8) {\tiny$9$};

\end{tikzpicture}
\caption{Possible $5\prescript{2}{}{S}_{1/2} \rightarrow 6\prescript{2}{}{P}_{3/2}$ $\sigma^-$ transitions of ${}^{87}\text{Rb}$, $F_g = 1$.\label{fig:schemesm87rb_6p32}}
\end{figure}

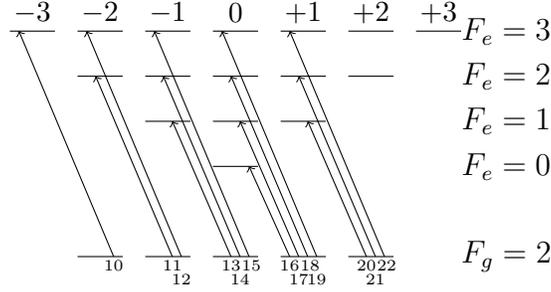
\begin{figure}[H]
\centering
 \begin{tikzpicture} [scale=0.6, decoration={coil,aspect=0.4,segment length=3mm,amplitude=3mm}]


\draw(1.5,1)--(2.5,1);
\draw(3,1) -- (4,1);
\draw(4.5,1) -- (5.5,1);
\draw(6,1) -- (7,1);
\draw(7.5,1)--(8.5,1);

\draw(4.5,3) -- (5.5,3);

\draw(3,4) -- (4,4);
\draw(4.5,4) -- (5.5,4);
\draw(6,4) -- (7,4);

\draw(1.5,5) -- (2.5,5);
\draw(3,5) -- (4,5);
\draw(4.5,5) -- (5.5,5);
\draw(6,5) -- (7,5);
\draw(7.5,5) -- (8.5,5);

\draw(0,6) -- (1,6);
\draw(1.5,6) -- (2.5,6);
\draw(3,6) -- (4,6);
\draw(4.5,6) -- (5.5,6);
\draw(6,6) -- (7,6);
\draw(7.5,6) -- (8.5,6);
\draw(9,6) -- (10,6);

\node at (11,1) {$F_g=2$};
\node at (11,3) {$F_e=0$};
\node at (11,4) {$F_e=1$};
\node at (11,5) {$F_e=2$};
\node at (11,6) {$F_e=3$};

\node at (0.5,6.4) {\small $-3$};
\node at (2,6.4) {\small $-2$};
\node at (3.5,6.4) {\small $-1$};
\node at (5,6.4) {\small $0$};
\node at (6.5,6.4) {\small $+1$};
\node at (8,6.4) {\small $+2$};
\node at (9.5,6.4) {\small $+3$};

\draw[->](3.8-1.5,1) -- (4.70-4.5,6);

\draw[->](3.8,1) -- (4.70-3,6);
\draw[->](3.6,1) -- (4.90-3,5);

\draw[->](3.8+1.5,1) -- (4.70-1.5,6);
\draw[->](3.6+1.5,1) -- (4.90-1.5,5);
\draw[->](3.4+1.5,1) -- (5.10-1.5,4);

\draw[->](3.8+3,1) -- (4.70,6);
\draw[->](3.6+3,1) -- (4.90,5);
\draw[->](3.4+3,1) -- (5.10,4);
\draw[->](3.2+3,1) -- (5.30,3);

\draw[->](3.8+4.5,1) -- (4.70+1.5,6);
\draw[->](3.6+4.5,1) -- (4.90+1.5,5);
\draw[->](3.4+4.5,1) -- (5.10+1.5,4);

\node at (3.8-1.5,0.8) {\tiny $10$};

\node at (3.6,0.8) {\tiny $11$};
\node at (3.8,0.5) {\tiny $12$};

\node at (3.4+1.5,0.8) {\tiny $13$};
\node at (3.6+1.5,0.5) {\tiny $14$};
\node at (3.8+1.55,0.8) {\tiny $15$};

\node at (3.2+1.5+1.5,0.8) {\tiny $16$};
\node at (3.4+1.5+1.5,0.5) {\tiny $17$};
\node at (3.6+1.5+1.55,0.8) {\tiny $18$};
\node at (3.8+1.5+1.5,0.5) {\tiny $19$};

\node at (3.4+4.5,0.8) {\tiny $20$};
\node at (3.6+4.5,0.5) {\tiny $21$};
\node at (3.8+4.55,0.8) {\tiny $22$};

\end{tikzpicture}
\caption{Possible $5\prescript{2}{}{S}_{1/2} \rightarrow 6\prescript{2}{}{P}_{3/2}$ $\sigma^-$ transitions of ${}^{87}\text{Rb}$, $F_g = 2$.\label{fig:2schemesm87rb_6p32}}
\end{figure}

\begin{figure}[H]
\centering
\includegraphics[scale=0.6]{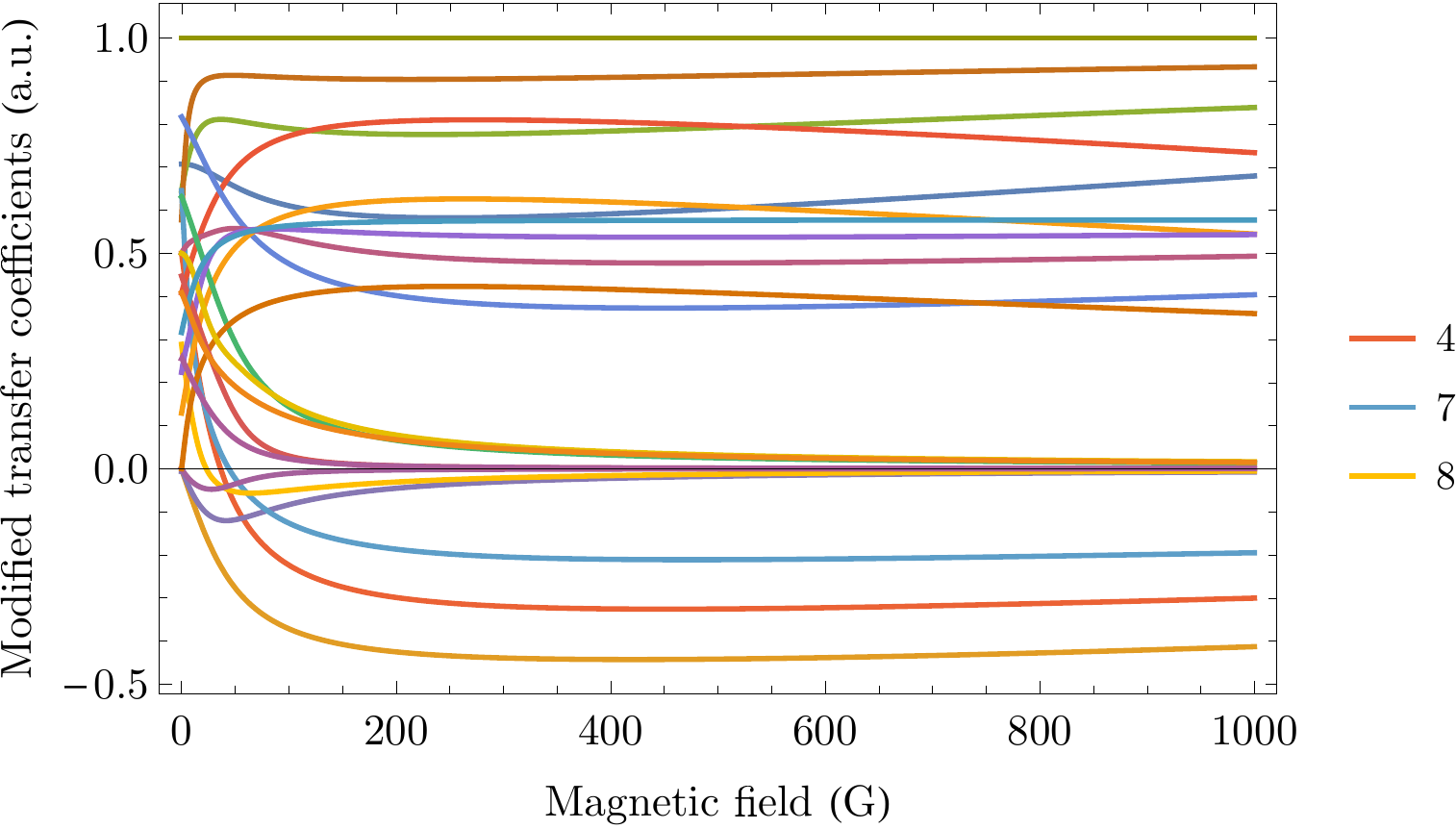}
\caption{$5\prescript{2}{}{S}_{1/2} \rightarrow 6\prescript{2}{}{P}_{3/2}$ $\sigma^-$ modified transfer coefficients. For the sake of clarity, only the ones with a cancellation are labelled. \label{fig:transcoef_87rb_smP32}}
\end{figure}

We schematize the possible $\sigma^-$ transitions on figure \ref{fig:schemesm87rb_6p32} and \ref{fig:2schemesm87rb_6p32}. As for the $\sigma^+$ case, $22$ transitions are possible in total. The modified transfer coefficients corresponding to all these transitions are represented on figure \ref{fig:transcoef_87rb_smP32}. Since cancelled transitions involve $3\times 3$ or $4\times 4$ blocks, we do not derive any analytical formula although it should be possible based on Ferrari and Cardano's formulas.

\begin{table}[hbtp]
\centering
\begin{tabular}{|c|c|c|c|}
\hline

Atom                                 & Transitions & $B$(G)      & $B^*$(G)         \\ \hline
                                     & $4$         & 36.32(16)  & 36.320508551(38) \\ \cline{2-4} 
                                     & $7$         & 44.13(31)  & 44.137763913(46) \\ \cline{2-4} 
\multirow{-3}{*}{${}^{87}\text{Rb}$} & $8$         & 23.01(18) & 23.016986486(24) \\ \hline
\end{tabular}
\caption{$B$-values cancelling ${}^{87}\text{Rb}$ $5\prescript{2}{}{S}_{1/2} \rightarrow 6\prescript{2}{}{P}_{3/2}$ $\sigma^-$ transitions.}
\label{tab:values87rbsmP32}
\end{table}

Again, as on figure \ref{fig:transcoef_87rb_spP32}, we observe a constant horizontal line corresponding to transition 10 which is a transition between pure states \linebreak ($m_{F_g}=-2
\rightarrow m_{F_e}=-3$). Among the other curves, three get cancelled (transitions $4$, $7$ and $8$, corresponding to respectively $m_{F_g} = 0,+1,+1$) for precise values of $B$. Unlike before, no transition starting from $F_g = 2$ is cancelled.
We will now present numerical data obtained for the all $5\prescript{2}{}{S}_{1/2} \rightarrow 6\prescript{2}{}{P}_{3/2}$ transition of ${}^{85}\text{Rb}$.

\section{$5\prescript{2}{}{S}_{1/2} \rightarrow 6\prescript{2}{}{P}_{3/2}$ transitions of rubidium 85}
Hereafter we present all the $5\prescript{2}{}{S}_{1/2} \rightarrow 6\prescript{2}{}{P}_{3/2}$ transitions of rubidium 85 which show a cancellation and their associated $B$-values. 
In this case, 116 transitions are possible in total, thus, for the sake of clarity, we will not include graphical representations.

\begin{table}[hbtp]
\centering
\begin{tabular}{|c|c|c|c|c|}
\hline

{ $F_g$} & { $F_e$} & { $m_{F_g}$} & { $B$(G)}      & { $B^*$(G)}          \\ \hline

{ $2$}   & { $3$}   & { $2$}       & { 8.759(18)}  & { 8.7592563674(69)}  \\ \hline

{ $2$}   & { $3$}   & { $1$}       & { 9.456(17)}  & { 9.4569409901(74)}  \\ \hline

{ $2$}   & { $3$}   & { $0$}       & { 10.393(14)} & { 10.3937170637(80)} \\ \hline

{ $2$}   & { $3$}   & { $-1$}      & { 14.972(15)}  & { 14.972413447(12)}  \\ \hline

{ $2$}   & { $2$}   & { $0$}       & { 16.224(22)}  & { 16.224059190(13)}  \\ \hline

{ $2$}   & { $2$}   & { $1$}       & { 16.708(28)}  & { 16.708656930(13)}  \\ \hline

{ $2$}   & { $2$}   & { $2$}       & { 16.853(33)}  & { 16.853016860(14)}  \\ \hline

{ $2$}   & { $3$}   & { $2$}       & { 4139(22)}    & { 4139.600083(11)}   \\ \hline

{ $2$}   & { $3$}   & { $1$}       & { 4467(22)}    & { 4467.802090(12)}   \\ \hline

{ $2$}   & { $3$}   & { $0$}       & { 4807(24)}     & { 4807.443620(13)}   \\ \hline

{ $2$}   & { $4$}   & { $-1$}      & { 5158(27)}     & { 5158.554086(13)}   \\ \hline
\end{tabular}
\caption{$B$-values cancelling ${}^{85}\text{Rb}$ $5\prescript{2}{}{S}_{1/2} \rightarrow 6\prescript{2}{}{P}_{3/2}$ $\sigma^-$ transitions.}
\label{tab:values85rbsmP32}
\end{table}

\begin{table}[H]
\centering
\begin{tabular}{|c|c|c|c|c|}
\hline

$F_g$ & $F_e$ & $m_{F_g}$ & $B$(G)                             & $B^*$(G)                                 \\ \hline

$2$   & $2$   & $-1$      & { 2.2199(46)}  & { 2.2199898843(17)}  \\ \hline

$3$   & $3$   & $-1$      & { 4.1621(39)}  & { 4.1621397468(32)}  \\ \hline

$3$   & $3$   & $-2$      & { 10.3043(87)} & { 10.3043976319(79)} \\ \hline

$3$   & $2$   & $2$       & { 10.953(13)}  & { 10.9539851230(87)} \\ \hline

$3$   & $2$   & $1$       & { 11.147(14)}  & { 11.1479877220(88)} \\ \hline

$3$   & $2$   & $0$       & { 11.334(14)}  & { 11.3343927931(88)} \\ \hline

$3$   & $2$   & $-1$      & { 11.416(14)}  & { 11.4168589034(88)} \\ \hline

$2$   & $3$   & $-2$      & { 14.322(16)}  & { 14.322575163(11)}  \\ \hline

$2$   & $3$   & $-1$      & { 14.860(17)}  & { 14.860250029(12)}  \\ \hline

$2$   & $3$   & $0$       & { 15.369(18)}  & { 15.369857179(12)}  \\ \hline

$2$   & $3$   & $-1$      & { 15.812(19)}  & { 15.812955622(12)}  \\ \hline

$2$   & $4$   & $-1$      & { 6271(30)}    & { 6271.265407(16)}   \\ \hline

$2$   & $4$   & $0$       & { 6603(33)}    & { 6603.849559(17)}   \\ \hline

$2$   & $4$   & $1$       & { 6942(36)}    & { 6942.545196(18)}   \\ \hline

$2$   & $4$   & $2$       & { 7285(40)}    & { 7285.070415(19)}   \\ \hline
\end{tabular}
\caption{$B$-values cancelling ${}^{85}\text{Rb}$ $5\prescript{2}{}{S}_{1/2} \rightarrow 6\prescript{2}{}{P}_{3/2}$ $\pi$ transitions.}
\label{tab:values85rbpiP32}
\end{table}

In table \ref{tab:values85rbpiP32}, all the $B$-values obtained are either between $2$ and $16$ G ie. small values of $B$, either after $6200$ G ie. huge values of $B$. For the last 4 transitions, the uncertainties are huge, similarly to transitions 
$1$, $10$ and $13$ of figure \ref{fig:transcoef_87rb_spP32}. 

\begin{table}[H]
\centering
\begin{tabular}{|c|c|c|c|c|}
\hline

{ $F_g$} & { $F_e$} & { $m_{F_g}$} & { $B$(G)}     & { $B^*$(G)}         \\ \hline

{ $3$}   & { $2$}   & { $1$}       & { 4.7932(69)} & { 4.7932030491(38)} \\ \hline

{ $3$}   & { $2$}   & { $0$}       & { 4.9968(73)} & { 4.9968609804(39)} \\ \hline

{ $3$}   & { $2$}   & { $-1$}      & { 5.2424(79)} & { 5.2424910523(40)} \\ \hline

{ $3$}   & { $2$}   & { $-2$}      & { 5.5486(85)} & { 5.5486439413(42)} \\ \hline

{ $3$}   & { $3$}   & { $-2$}      & { 20.228(20)} & { 20.228037283(16)} \\ \hline

{ $3$}   & { $3$}   & { $-1$}      & { 23.044(21)} & { 23.044654730(18)} \\ \hline

{ $3$}   & { $3$}   & { $0$}       & { 26.036(21)} & { 26.036555825(21)} \\ \hline

{ $3$}   & { $3$}   & { $1$}       & { 29.242(21)} & { 29.242109204(24)} \\ \hline

{ $3$}   & { $3$}   & { $2$}       & { 32.693(22)} & { 32.693414753(27)} \\ \hline

{ $2$}   & { $1$}   & { $0$}       & { 50.02(54)}  & { 50.018924381(39)} \\ \hline

{ $2$}   & { $1$}   & { $-1$}      & { 56.18(53)}  & { 56.182412663(44)} \\ \hline

{ $2$}   & { $1$}   & { $-2$}      & { 63.40(50)}  & { 63.398936275(50)} \\ \hline

{ $3$}   & { $1$}   & { $0$}       & { 72.46(47)}  & { 72.459280144(56)} \\ \hline

{ $3$}   & { $1$}   & { $-1$}      & { 79.54(49)}  & { 79.538277943(62)} \\ \hline

{ $3$}   & { $1$}   & { $-2$}      & { 87.42(52)}  & { 87.425519697(67)} \\ \hline

{ $3$}   & { $2$}   & { $-3$}      & { 96.22(53)}  & { 96.220676416(72)} \\ \hline
\end{tabular}
\caption{$B$-values cancelling ${}^{85}\text{Rb}$ $5\prescript{2}{}{S}_{1/2} \rightarrow 6\prescript{2}{}{P}_{3/2}$ $\sigma^+$ transitions.}
\label{tab:values85rbspP32}
\end{table}

\section{Conclusion and perspectives}

In this paper, we dressed an overall view of the theoretical behavior of $5S \rightarrow 6P$ transition cancellations for both Rubidium isotopes. To sum up, all the constants involved are known to around $10^{-10}$ which is not the case for the ESED. The experimental measurements is envisaged to be done soon at the Institute for Physical Research of Ashtarak, Armenia. To obtain precise results and avoid transition overlapping due to Doppler broadening, sub-Doppler methods have to be used here. This could be achieved using saturated absorption, in which sub-Doppler resolution is attained by forming atomic velocity-selective optical pumping (VSOP) resonances. These VSOPs are accompanied by strong crossover (CO) resonances which complicate the spectra \cite{klinger}. Another way to obtain sub-Doppler resolution is to use a nano-cell, since Doppler broadening is simply annihilated by the geometry of the cell. Considerably good results have been obtained using nanocells \cite{sargsyan1,hrant2} in which the magnetic field can be considered to be uniform due to the cell's thinness. 
Experimental results will be highly influenced by the experimental technique that will be used (or developed) to measure the measure the magnetic field. In every table, we showed the $B$-values computed when taking into account the uncertainty of every parameter, and the $B^*$-values computed when assuming the ESED to be exact. It is clearly seen that the precision on these values dramatically decreases when the uncertainties on the ESED are taken into account. 
The measured values could be used to refine uncertainties of all the parameters involved in the problems: ESED, Bohr magneton and even Landé factors. 
In a reverse way, making an experiment to refine the values of the ESED would also increase the precision of our computations. Moreover, the $B$-values are determined by analytical formulas meaning they could be used as standard to calibrate magnetometers.

\section{Funding Information}

A. Aleksanyan thanks the Graduate School EUR EIPHI for the funding CO.17049.PAC.AN.

\bibliographystyle{elsarticle-num-names} 
\bibliography{biblio}


\end{document}